\documentclass{article}
\usepackage[margin=1in]{geometry}

\usepackage[section]{algorithm}      
\usepackage[noend]{algpseudocode}    
\usepackage{amssymb}
\usepackage{subcaption}

\usepackage{amsmath}
\usepackage{graphicx}
\usepackage[colorinlistoftodos]{todonotes}
\usepackage{multirow}
\usepackage{booktabs}
\usepackage{caption}
\usepackage{tabularx}
\usepackage{xcolor}
\usepackage{gensymb}
\usepackage{float}
\usepackage[capitalize,nameinlink,noabbrev]{cleveref} 
\usepackage{gensymb} 
\usepackage{algorithm}    
\usepackage{placeins}
\usepackage{algpseudocode}    

\usepackage{tikz}
\usetikzlibrary{shapes,arrows,positioning}

\DeclareMathSizes{10}{9.5}{6}{6}
\makeatletter
\g@addto@macro \normalsize {%
 \setlength\abovedisplayskip{5pt plus 2pt minus 2pt}%
 \setlength\belowdisplayskip{5pt plus 2pt minus 2pt}%
}
\makeatother

\newcommand{\Exp}{\mathbb{E}} 
\newcommand{\Var}{\mathbb{V}} 
\newcommand{\Cov}{\mathrm{Cov}} 
\newcommand{\bigO}{\mathcal{O}} 
\newcommand{\esssup}{\mathrm{sup\,ess}}
\newcommand{\essinf}{\mathrm{inf\,ess}}
\newcommand{\RE}{\mathcal{R}} 

\newcommand{\unk}{\mathbf{u}}  
\newcommand{\datan}{y} 
\newcommand{\data}{\mathbf{\datan}} 
\newcommand{\mpr}{p_0}         
\newcommand{\llh}{L}           
\newcommand{\mps}{p_{post}}   
\newcommand{\fwd}{\mathcal{G}} 
\newcommand{\ns}{\eta}         
\newcommand{\mns}{p_\ns}      
\newcommand{\mob}{p_l}         
\newcommand{\prop}{\tilde{\unk}} 

\newcommand{\fim}{\mathcal{I}}
\newcommand{\prm}{\theta}
\newcommand{\mpsp}{\mps^{(\prm)}}
\newcommand{\mprp}{\mpr^{(\prm)}}
\newcommand{\mobp}{\mob^{(\prm)}}
\newcommand{\zp}{Z^{(\prm)}}
\newcommand{\pot}{\Phi}
\newcommand{\potp}{\pot^{(\prm)}}

\newcommand{\ft}{\tilde{f}}
\newcommand{\xt}{\tilde{x}}
\newcommand{\ub}{U}

\newcommand{\agn}{w}
\newcommand{\ag}{\mathbf{\agn}} 

\newcommand{\erf}{\text{erf}}

\newcommand{\gr}{\hat{R}}

\begin{document}
\title{Embracing Uncertainty in ``Small Data'' Problems: Estimating Earthquakes from Historical Anecdotes}
 \author{N. E. Glatt-Holtz, R. A. Harris, A. J. Holbrook, J. Krometis,\\ Y. Kurniawan, H. Ringer, \& J. P. Whitehead\\
 \scriptsize{emails: negh@iu.edu,rharris@byu.edu,aholbroo@g.ucla.edu,jkrometis@vt.edu,}\\ \scriptsize{yonatank@student.byu.edu,hringer@vt.edu, whitehead@mathematics.byu.edu}}
 \date{\today}



\maketitle

\begin{abstract}
Seismic risk estimates will be vastly improved with an increased understanding of historical (and pre-historical) seismic events.
However the only existing data for these events is anecdotal and sparse.  To address this we developed a framework based on Bayesian inference to estimate the location and magnitude of pre-instrumental earthquakes.  We present a careful analysis of results obtained from this procedure which justifies the sampling algorithm, its convergence to the resultant posterior distribution, and yields estimates on uncertainties in the relevant quantities.  Using a priori estimates on the posterior and numerical approximations of the Hessian, we demonstrate that the 1852 Banda Sea earthquake and tsunami is indeed well-understood given certain explicit hypotheses.  Using the same techniques we also find that the 1820 south Sulawesi event may best be explained by a dual fault rupture, best attributed to the Kalatoa fault potentially conjoining the Flores thrust and Walanae/Selayar fault.
\end{abstract}

\section{Introduction}

The geologically recent mega-thrust earthquakes and giant tsunamis in Indonesia (2004) and Japan (2011) as well as the seismic catastrophes in Haiti (2010) and China (2008) all occurred in regions previously mapped as having ``low'' seismic hazard. One reason for this is because hazard assessments relied largely on instrumental data only available since the mid-1900s \cite{Steinetal2012}. Historical records and geological evidence of seismic events exist in each of these regions, but they were not adequately accounted for due to the uncertainty that is inherent to such data sources \cite{MuJi2008}. For example, tsunami deposits were documented on the Sendai Plain before the 2011 Japan mega-thrust earthquake \cite{minoura2001869}, but were not considered in risk assessments, such as the retrofitting of the nuclear power plant.  These recent seismic and tsunami disasters motivate us to push beyond the geologically limited time window of instrumentally recorded earthquakes to find new ways of quantifying unconventional data sources for earthquake locations and magnitudes.

Recent advances in sensing technology and rapid progress in data management and modeling has made it so that `big data' is an essential part of modern seismology \cite{arrowsmith2022big} as well as many other sub-fields of geophysics \cite{yu2021deep}.  Yet there is no escaping the fact that the temporal scale of seismic activity far exceeds the resolution of modern instruments. The rub of the matter is that `big data' methods are incapable of breaching the gap in temporal scale for seismology precisely because the data of interest is sparse, irregular, and far to uncertain for `big data' techniques.
This article reports on one approach to this `small data problem' which uses modern Bayesian statistical techniques and computational resources to ascertain the probable location and magnitude of pre-Cold War earthquakes in the Indonesian archipelago solely using anecdotal records from Dutch colonists (see \cite{ringer2021methodological,paskett2024tale}).  We demonstrate that the previously published results using this method are in fact statistically robust, and under the hypotheses proposed, provide an unbiased estimate of the posterior distribution of the causal earthquake.

Resilience to tsunamis and other seismic hazards requires learning from what has happened in the geological past
\cite{unisdr2009}. Evidence of previous earthquake and tsunami events can be found in the geological record, such as deposits left by previous tsunami events \cite{Sulaemanetal2017}, damage to coral reefs \cite{meltzner2010coral,gagan2015coral}, and sediment cores of turbidites \cite{GoNeJo2003}. Additional information on past events is also available in the form of textual accounts in historical records such as the Wichmann catalog \cite{wichmann1918earthquakes,wichmann1922earthquakes,harris2016waves} and other sources \cite{Re1858,Be1868,Mu2012,Re2012} for Indonesia (see \cite{martin2022gempa} for a comprehensive catalog for the Indonesian archipelago), North and South America (see \cite{SiAsHa1981,De2004,KoKo2004} for example), and many other locations in Asia and throughout the world. For this data we have seen a shift from the question ``can we quantify what happened?'' to ``can we make a principled estimate of the uncertainties around what happened?''.

Previous efforts to reconstruct pre-instrumental earthquakes have varied from a focus on the use of geological evidence (see \cite{monecke20081,sieh2008earthquake,jankaew2008medieval,meltzner2010coral} for example) to the use of historically recorded (but not instrumental) accounts \cite{okal2003mechanism,bryant2007cosmogenic,LiuHarris2014,harris2016waves,reid2016two,fisherharris2016,GrNgCuCi2018,Cummins2020,PranantyoCummins2020} as well as some combination of the two types of uncertain data (see \cite{martin2019reassessment} for one example).  Most of these efforts, particularly those directed toward using historical records, have relied on a combination of physical intuition and a restricted number of forward simulations to match the observational data.  Qualitative comparisons are then made to the historical (or geological) record, and a heuristic choice is made as to the ``best'' forward simulation that fits the data.

In this paper, we analyze the problem of reconstructing historical earthquakes from accounts of tsunamigenic impact using a Bayesian framework \cite{tarantola2005inverse,kaipio2005statistical,stuart2010inverse,gelman2014bayesian}. The Bayesian perspective is a natural fit because the chief problem we face is uncertainty in the data; the resulting posterior distribution will therefore provide estimates of the most likely values of, but also the uncertainties that surround, the seismic parameters we would like to estimate, e.g., the magnitude and location of the historical earthquake in question.
Here the numerical resolution of partial differential equations (PDEs) describing tsunami wave propagation provides a ``forward map'' which can be ``inverted'' to adequately estimate
the posterior distribution.
While the Bayesian framework has been used in the past to address problems in seismicity (see \cite{bui2013computational,martin2012stochastic,giraldi2017bayesian} for a few examples), the approach first outlined in \cite{ringer2021methodological} and further used in \cite{paskett2024tale,wonnacott2024methodological} is the first that we know of to apply this methodology to pre-instrumental seismic events.  The purpose of this article is to summarize this approach, and to present statistically validating estimates that verify the accuracy of this reconstruction and the consequent posterior distribution.

\section{The Data: Historical Accounts of Tsunamis} \label{sec:data}
Although there is a significant amount of geological evidence of tsunamis and earthquakes (see \cite{GoNeJo2003,meltzner2010coral,gagan2015coral,Sulaemanetal2017,martin2022gempa} for example), we will focus here on historical records of tsunami impact alone, as the uncertainties in such observations, while significant relative to modern seismometers, are significantly less than geological data or historical accounts of earth shaking.  Very recently, the catalog Gempa Nusantara \cite{martin2022gempa} was published which includes references to historical accounts of tsunami and earthquake events throughout the Indonesian archipelago.  This catalog coupled with Arthur Wichmann's \emph{The Earthquakes of the Indian Archipelago} \cite{wichmann1918earthquakes,wichmann1922earthquakes,harris2016waves} provides as accurate a documentation as can be expected for these historical events.  The two events focused on in this report are drawn primarily from the Wichmann catalog (see \cite{ringer2021methodological,paskett2024tale} for more details on each particular event).  As an explicit example, consider the textual account below which describes the impact of a tsunami at Banda Neira island located in the center of the Banda Sea.

\fbox{\begin{minipage}{35em}
{\bf 1852, November 26, 7:40.} At Banda Neira, barely had the ground been calm for a quarter of an hour when the flood wave crashed in \dots The water rose to the roofs of the storehouses \dots and reached the base of the hill on which Fort Belgica is built on.
\end{minipage}}

This excerpt from the Wichmann catalog provides clear descriptors, such as the location, arrival time, wave height, and inundation length that can be used to characterize the tsunami and infer the earthquake that might have caused it. Moreover, because historical observations of the tsunamis were observed in multiple, often geographically-dispersed locations (Banda Neira being just one of several for the 1852 event), even uncertain observations can be ``triangulated'' to provide more certain estimates of earthquake size and location. 

At the same time, the excerpt also demonstrates some of the challenges associated with doing such an inference in a rigorous way: Given that these measurements were taken well before the modern era of automated and sophisticated sensing, how accurate are they? What does water rising to rooftops tell us about the event?

\section{Methods}\label{sec:methods}
\subsection{Bayesian Inference} \label{sec:bayes}
Our approach to leveraging the data described in \cref{sec:data} in a more principled and systematic fashion uses a Bayesian framework \cite{gelman2014bayesian,kaipio2005statistical,dashti2017bayesian,tarantola2005inverse}, which provides a rigorous, statistical methodology for converting uncertain outputs into probabilistic estimates of model parameters. While the Bayesian approach is not novel, and is in fact in use throughout the Geosciences more generally, we elaborate its application in this setting in great detail to highlight the novel approach introduced in \cite{ringer2021methodological}.

We note that while we use the pre-instrumental earthquake-tsunami problem as a motivating example, the framework described in this section is applicable to any problem where the ``data'' is ``small'' i.e. sparse and/or highly uncertain.
In what follows, we denote the model parameters characterizing the seismic event of interest by $\unk$, the ``data'' by $\data$, the prior measure by $\mpr$, the forward model $\fwd$ from model parameters (e.g., earthquake magnitude and location) to observables (e.g., tsunami wave height), the likelihood by $\llh(\unk;\data) \propto p(\data | \unk)$, and the posterior measure by $\mps$. See the references above for definitions of these quantities.

Bayes' Theorem provides an explicit expression for $\mps$ as
\begin{align}
  \label{eq:bayes}
  \mps(\unk)
  \propto \llh(\unk;\data) \mpr(\unk).
\end{align}
Most critically, the Bayesian approach incorporates uncertainty at all levels of the inverse problem, an essential feature given that the data in this case clearly does not provide enough information to fully specify the model parameters -- we hope that it will tell us \emph{something} about the parameters, but expect that it will necessarily not tell us \emph{everything}.

\subsection{Likelihood Modeling} \label{sec:llh}

In this section, we outline our procedure for modeling noisy or anecdotal data via the likelihood. While the tsunami observations described in \cref{sec:data} provide a motivating example, the approach described here could be applied to any problem where the data has similar issues of being so ill-defined as to be anecdotal. Specific application to pre-instrumental seismic events in Indonesia are detailed in \cite{ringer2021methodological,paskett2024tale,wonnacott2024methodological}.

To model the data $\data$ from anecdotal observations like those described in \cref{sec:data}, we adopt a data augmentation approach (see, e.g., \cite{tanner1987calculation,albert1993bayesian,van2001art,holbrook2022bayesian}) and introduce an auxiliary variable $\ag$ representing additional, unobserved data. 
Then the likelihood is given by
\begin{align}\label{eq:llh:aug}
    \llh(\unk;\data) 
    &\propto p(\data|\unk) = \int p(\data|\ag,\unk) p(\ag|\unk) \, d\ag.
\end{align}
Here, for our augmentation variable $\ag$ we use the true value of the output (e.g., $\agn_i$ might be the true wave height while $\datan_i$ is the observed value of the wave height) and assume that uncertainty in the true value is independent of $\unk$ so that $p(\data|\ag,\unk)=p(\data|\ag)$.
In this context, we incorporate information from the anecdotal data by directly modeling 
$p(\data|\ag)$ as a function of $\ag$ via a fixed probability density function $\mob(\ag)$ and assume
\begin{align} \label{eq:mob}
  \mob(\ag) \propto  p(\data|\ag) .
\end{align}
From a practical standpoint, this strategy leverages the fact that the unexpressed integration constant in \eqref{eq:mob} does not feature in the posterior distribution \eqref{eq:bayes}:
\begin{align*}
      \mps(\unk)
  = \frac{ \left( \int p(\data|\ag) p(\ag|\unk) \, d\ag  \right)\mpr(\unk)}{\int \left( \int p(\data|\ag) p(\ag|\unk) \, d\ag  \right)\mpr(\unk) d\unk} = \frac{ \left( \int \mob(\ag)  p(\ag|\unk) \, d\ag  \right)\mpr(\unk)}{\int \left( \int \mob(\ag)  p(\ag|\unk) \, d\ag  \right)\mpr(\unk) d\unk} \, .
\end{align*}
From a philosophical standpoint, this strategy represents an application of the likelihood principle \cite{fisher1922mathematical,casella2021statistical}, insofar as it allocates higher values of $\mob(\cdot)$ for $\ag$ that better `match' the data as quantified by the function $p(\data|\ag)$.

Meanwhile, we require that true observations $\ag$ match the forward map $\fwd(\unk)$ (in our setting the forward map is implemented via the Geoclaw software package \cite{leveque2011tsunami}), so we define
\begin{align} \label{eq:dirac}
    p(\ag|\unk) &= \delta(\ag-\fwd(\unk)),
\end{align}
where $\delta$ is the Dirac distribution centered at zero. 
Then plugging \eqref{eq:mob} and \eqref{eq:dirac} into \eqref{eq:llh:aug} yields
\begin{align}\label{eq:llh}
    \llh(\unk;\data) 
    &\propto \int \mob(\ag) \delta(\ag-\fwd(\unk)) \, d\ag
    = \mob(\fwd(\unk)).
\end{align}
 
To compare this with a more standard representation, assume the data model $\data = \ag+\ns$ for additive observational noise where $\ns \perp \unk$ and $\ag = \fwd(\unk)$ for some $\unk$. Denoting the probability density of $\ns$ by $\mns$, we see that $\ag$ has density $\mob(\ag):=\mns(\data-\ag)$ and via \eqref{eq:llh},
\begin{align}
    \label{eq:llh:add}
    \llh(\unk; \data) 
    = \mns(\data-\fwd(\unk)).
\end{align}
This is precisely a  popular likelihood used in Bayesian inverse problems; see, e.g., \cite[Section 3.2.1]{kaipio2005statistical} or \cite[Section 1.1]{dashti2017bayesian}. Hence, \eqref{eq:mob} is a generalization fo the standard likelihood model where the structure of the observational noise is left in a more implicit form, i.e. it is not necessarily additive nor orthogonal to the input of the model.  This represents a fundamental shift in the traditional Bayesian inversion approach that can incorporate both the epistemic (reducible) and aleatoric (irreducible)  uncertainties in the problem.  To be more precise, uncertainty in the observables of interest in this setting are inherently aleatoric, there is no way to improve the accuracy of the observable data, and yet by deviating from the assumption of additive noise, we can account for this uncertainty as well as the epistemic uncertainty that for example arises from modeling assumptions.

Of course, the choice of the observation distribution $\mob$ in \eqref{eq:mob} is subjective, as any interpretation of the historical records described in \cref{sec:data} must be. However, the approach outlined above represents a clear improvement over the modeling of the historical data as a single numerical value in at least two ways: 
\begin{itemize}
    \item By using probability distributions rather than single values, the methodology more clearly encapsulates the aleatoric uncertainty associated with the observations.
    \item Modeling assumptions are explicitly specified and incorporated into the methodology so that the results are rigorous and reproducible.
\end{itemize}
One might interpret the direct modeling of the likelihood distribution as repeating a deterministic approach to the inverse problem a large number of times, with the observation distribution $\mob(\ag)$ representing the probability that a given modeler might interpret the observation as representing the true value $\ag$. As pointed out above, this interpretation allows us to place estimates on the aleatoric uncertainty of the problem.

In any case this represents a fruitful paradigm shift from the usual Bayesian inversion framework by allowing more direct application to problems where observational signals and noise are inextricably intertwined, i.e. the aleatoric uncertainty is dominating. A practitioner can simply model what the observations tell them via $\mob$ in \eqref{eq:mob} and then proceed with the usual Bayesian inference using the likelihood in \eqref{eq:llh}. A direct extension of the current work would be to implement this approach for other types of geological evidence such as coral uplift, sediment cores, and disrupted turbidites, but the overall framework can be leveraged by problems outside of seismic inversion as well.  From a more methodological perspective, a comparison of this modern Bayesian approach (handling aleatoric uncertainty) with the Data Consistent Inversion method \cite{butler2014measure,butler2018combining,butler2020we,butler2020data,del2024sequential} which has been shown to adequately handle both types of uncertainty, would be very fruitful.

The full likelihood for the tsunami-earthquake inversion problem is complete with the specification of the forward model.  The forward model in this case maps the earthquake parameters which model the resulting seafloor deformation via the Okada model \cite{okada1985surface}, and then simulates the generation and propagation of the resultant tsunami to produce wave arrival times, wave heights, and inundation lengths at the observation locations. This is accomplished using the GeoClaw software package \cite{leveque2008high, leveque2011tsunami, gonzalez2011validation, berger2011geoclaw}. 

\subsection{A prior distribution, and posterior sampling via Markov Chain Monte Carlo} \label{sec:1852mcmc}

The prior distribution for the historical seismic investigation considered here is a probability distribution that describes the probable earthquake parameters that are physically feasible for the event in question.  For the 1852 Banda Sea earthquake, the latitude-longitude location, depth, and rough geometry of the underlying fault were specified using data from the Slab2 dataset \cite{HayesSlab2} which incorporates modern instrumental data to map out major subduction zones globally. No such dataset was available for either the Flores thrust nor Walanae/Selayar fault so that the prior distribution on these parameters for the 1820 Sulawesi event was specified using a mixture of Gaussian Processes and expert-informed geophysical insight.  The prior distribution on magnitude for both events was taken from the Gutenberg-Richter distribution, truncated to reasonable maximum (9.5) and minimum (6.5) values. The prior on the length, width and slip of the earthquake were taken as fluctuations around a mean fitted value determined by magnitude, and dictated by historical earthquake data cataloged in \cite{wells1994new} and from recent major events from the global USGS dataset.

The outcome of Bayesian inference is the posterior probability distribution. Computing this distribution in practice requires computing the normalization constant $Z$ in \eqref{eq:bayes}, an integral that can be difficult if not impossible to evaluate in practice. Instead we drew samples from the posterior distribution using Markov Chain Monte Carlo (MCMC) methods. Because we did not have an adjoint solver for this PDE-based forward map, gradient-based methods like Hamiltonian Monte Carlo were not available. We therefore employed random walk-style Metropolis-Hastings MCMC; a diagonal covariance structure was used for the proposal kernel with the step size in each of the selected earthquake (or landslide) parameters (see \cite{ringer2021methodological,paskett2024tale,wonnacott2024methodological} for details) was tuned to approximate the optimal acceptance rate of roughly $0.23$ \cite[Section 12.2]{gelman2014bayesian}.

The final standard deviations for the random walk proposal kernel are given in 
the GitHub repository at \url{http://github.com/jpw37/tsunamibayes}. Chains, particularly when initialized in different regions of the parameter space, sometimes got stuck in places with low posterior probability. We therefore conducted periodic importance-style resampling according to posterior probability (see \cite{doucet2001sequential}); this resampling does not maintain invariance with respect to the posterior measure, but provides a mechanism to ``jump'' trapped samples from poorly-performing regions of the parameter space to regions given more weight by the posterior distribution. To minimize any bias from the resampling steps, the approximate posterior was ultimately assembled from samples collected after a suitable burn in period following the last resampling step. The resulting algorithm is summarized in \cref{alg:mcmc}.

\begin{algorithm}[ht]
\caption{(MCMC as implemented for the pre-instrumentation inverse seismic problem)}\label{alg:mcmc}
\begin{algorithmic}[1]
   \State Choose number of chains $M$, resampling rate $N$, proposal covariance $C$, and initial parameters $\unk_0^{(i)}, i=1,\dots,M$.
    \For{$k \geq 0$}
        \For{$i = 1,\dots,M$}
            \State Propose $\prop^{(i)} = \unk_k^{(i)} + \eta, \eta \sim N(0,C)$
            \State Run Geoclaw to compute likelihood $\llh(\prop^{(i)};\data)$ according to \eqref{eq:llh} and \eqref{eq:1852llh}.
            \State Compute un-normalized posterior $\mps(\prop^{(i)})$ from \eqref{eq:bayes}.
            \State Set $\unk_{k+1}^{(i)} :=\prop^{(i)}$ with probability                $\min\{1,\mps(\prop^{(i)})/\mps(\unk_k^{(i)})\}$. 
            \State Otherwise take $\unk_{k+1}^{(i)} := \unk_{k}^{(i)}$.
        \EndFor
        \State If $k \mod N=0$, resample $\unk_k^{(i)} \sim \Sigma_j \mps(\unk_k^{(j)})\delta\left( \unk_k^{(j)} \right) /\Sigma_l \mps(\unk_k^{(l)}), i=1,\dots,M$.
        \State $k \to k + 1$.
    \EndFor
\end{algorithmic}
\end{algorithm}

\subsection{Verification of resolution of the posterior}
To verify that the posterior is adequately resolved, we focus on its geometry near the maximum a posteriori (MAP) value.  If the distribution is adequately resolved then the MAP point will be a maximum indicating that the gradient of the posterior should be zero, and as long as the distribution is sufficiently smooth then the Hessian (matrix of second order derivatives) will be negative definite.  As gradients are not available for the forward model $\fwd$ then we resort to surrogates or approximations of the derivative of the full posterior.  This is done via the method outlined here.

Let $f$ be a function whose input is $\boldsymbol{\theta} = \begin{pmatrix} \theta_1 & \theta_2 & \dots & \theta_N \end{pmatrix}^T$.
Suppose there are $M$ samples of $\boldsymbol{\theta}$, that is we have $M$ different possible inputs to $f$.  Denote $f^m = f(\boldsymbol{\theta}^m)$ as the model output evaluated using the parameter sample $\boldsymbol{\theta}^m = \begin{pmatrix} \theta_1^m & \theta_2^m & \dots & \theta_N^m \end{pmatrix}^T$.
The Taylor series expansion of $f^m$ around the specific set of inputs/parameters $\tilde{\boldsymbol{\theta}}$ is given by
\begin{equation}
    \label{eq:linear_taylor_fm}
    f^m \approx f(\tilde{\boldsymbol{\theta}}) + \sum_{n=1}^N \frac{\partial f (\tilde{\boldsymbol{\theta}})}{\partial \theta_n} (\theta_n^m - \tilde{\theta}_n).
\end{equation}
If we have $M$ samples of $\boldsymbol{\theta}$ and correpsondingly of $f$, then Eq.~\eqref{eq:linear_taylor_fm} defines a set of $M$ equations.
These equations can be written compactly in matrix notation as
\begin{equation}
    \label{eq:linear_taylor_fm_mat}
    \boldsymbol{F} \approx \Phi \boldsymbol{d},
\end{equation}
where the entries of the vector $F$ and matrix $\Phi$ are given by
\begin{gather*}
    F_m = f^m - f(\tilde{\boldsymbol{\theta}}), \\
    \Phi_{mn} = \theta_n^m - \tilde{\theta}_n, \text{ and } \\
    d_n = \frac{\partial f(\tilde{\boldsymbol{\theta}})}{\partial\theta_n}.
\end{gather*}
The gradient of the function $f$ at $\tilde{\boldsymbol{\theta}}$ can be approximated by solving Eq.~\eqref{eq:linear_taylor_fm_mat} for the vector $\boldsymbol{d}$.

We can extend Eq.~\eqref{eq:linear_taylor_fm} to second order to yield a similar approach to estimate the second order derivatives as well, requiring an even larger number of samples $M$ in order to yield a reasonable estimate of the resultant Hessian matrix.  In this case the unknown vector will now include elements from both the Jacobian (gradient if $f(\boldsymbol{\theta})$ is scalar valued) and the Hessian matrices.
Solving this new linear problem yields an estimation of both the Jacobian and the Hessian which can be used to verify whether the MAP is adequately realized.

For the posterior distributions considered here, the parameters have different physical units (length of the earthquake rupture is in thousands of kilometers whereas the slip length is in meters for example), and thus they have a significantly different range of values meaning that the estimated derivative will be unnecessarily skewed if computed in the original physical units.
To avoid this potential pitfall when estimating the derivative, we scale the sample parameters $\boldsymbol{\theta}$ from the posterior distribution (as sampled via MCMC) by the standard deviation across the entire sampled posterior, i.e. each different earthquake parameter is rescaled according to its variance over the entire sampled posterior distribution.  Additionally, in order to obtain a good estimation of the derivative, we want to use samples $\theta^m$ that are not too far from the estimation point of interest $\tilde{\boldsymbol{\theta}}$ in the estimation, i.e., the MAP.
To identify how `close' we need to draw the samples to the MAP, we define a hyper-sphere of radius $r$ in the scaled parameter space around the MAP point and only use the parameter samples that are located inside this hyper-sphere, which provides the hyperparameter $r$ which can be selected to give the most reasonable derivative approximation.
Large values of $r$ includes samples that are far from the MAP which will skew the derivative approximation, while small values of $r$ gives too few samples in the regression, in either extreme the derivative estimation will be inaccurate.
We choose an $r$ value that minimizes the norm of the gradient of the posterior at the MAP, i.e. we are relying on the assumption that the MAP truly is a local maximum of the posterior, i.e a critical point.

\subsection{Error Bounds and Sensitivity of the Posterior Distribution} \label{sec:errorbounds} 

Given the necessarily uncertain process of interpreting textual records as probability distributions, it would be natural to question how sensitive the results are to different choices of the observation distribution. 
To allay some of these concerns, in this section we describe the computation of upper and lower bounds on the potential error in the posterior distribution due to the choice of likelihood, specifically the selected parameterization of the observational probability distributions. In other words, we produce numerically estimated bounds (based on rigorously justified assumptions) that justify that small perturbations in our choice of the observational probability distributions and/or parameterizations of the prior, will not drastically affect the structure of the resultant posterior distribution, i.e. the posterior itself is \emph{not} overly sensitive to the assumptions made in the setup described here and hence we get a reasonable grasp on both the aleatoric and epistemic uncertainty of the problem.

To develop these bounds, we use three theoretical results from \cite{dupuis2016path}, which provides various bounds on estimates derived from probability distributions as those distributions are perturbed. First, we use a second order estimate for the relative entropy (Kullback–Leibler divergence) $\RE$ between a distribution $P^{\prm}$ and its perturbation $P^{\prm+v}$ (see \cite[Equation 2.35]{dupuis2016path}):
\begin{align}\label{eq:dupuis235}
    \RE (P^{\prm+v} || P^{\prm} ) = \frac{1}{2} v^T \fim\left( P^\prm \right)v + \bigO\left( |v|^3 \right),
\end{align}
where $\fim$ is the Fisher information matrix (FIM). The $\fim$ that we specify here corresponds to a measurement of the dependence of information contained in the posterior, on the specific parameters selected to create the observational probability distributions which then define the likelihood (when coupled with the forward model). 

The second estimate utilizes the following bound on the expected value of an observable $f$ (an observable in this context is any variable continuously drawn from the posterior, i.e. the resultant earthquake parameters are one example) with respect to the probability measure $Q$ in terms of the values of $f$ according to the probability measure $P$ \cite[Equation 2.11]{dupuis2016path}:
\begin{equation}\label{eq:dupuis211}
\begin{aligned}
\sup_{c>0} &\left( -\frac{1}{c} \log \Exp_{P} \left[e^{-c(f -\Exp_{P} [f])}\right] - \frac{1}{c} \RE (Q || P ) \right)\\
    &\qquad \le
    \Exp_{Q}[f] - \Exp_{P}[f] \le \inf_{c>0} \left( \frac{1}{c} \log \Exp_{P}\left[e^{c(f -\Exp_{P} [f])}\right] + \frac{1}{c} \RE (Q || P ) \right).
\end{aligned}
\end{equation}
Here $\Exp_{P}$ and $\Exp_{Q}$ are the expected values according to models $P$ and $Q$, respectively, and $Q$ is assumed to be absolutely continuous with respect to $P$ (that is, $P(A)=0 \implies Q(A)=0$ for any event $A$). By letting $P$ be the posterior measure, we can estimate the uncertainty in observables with respect to other probability measures $Q$ which are `close' to the estimated posterior $P$, i.e. we can estimate how much slight changes in the posterior would effect the observables we care about.

The expressions inside the $\sup$ and $\inf$ in \eqref{eq:dupuis211} can be differentiated to find the value of $c$ that provides the optimal bound for a given value of $\RE (Q || P )$. When $f$ is bounded, the equations also give a global upper and lower bound $\Exp_Q[f]$. Details of these derivations are reserved for \ref{app:dupuis211}. 
By combining these bounds with \eqref{eq:dupuis235} (setting $Q = P^{\prm+v}$), we can estimate bounds on our estimates of earthquake parameters for perturbations of the observation distributions. To model the worst-case scenario (most sensitive to changes in the parameters), we assume a perturbation in the direction of the first singular vector of the Fisher Information Matrix $\fim$.

Finally, \cite[Equation 2.39]{dupuis2016path} gives bounds on the sensitivity of estimates of observables due to changes in the likelihood (more directly changes to the observational probability distributions, i.e. perturbations in $\prm$):
\begin{align}\label{eq:dupuis239}
    \left| S_{f,v}\left( P^\prm \right) \right| \le \sqrt{ \Var_{P^\prm}(f) } \sqrt{ v^T \fim\left( P^\prm \right)v }.
\end{align}
Here $f$ is an observable, which we will consider to be the selected earthquake parameters that are used to model the earthquake and hence tsunami (typically we use a variation of the Okada parameters \cite{okada1985surface}), $\Var$ denotes the variance, and the sensitivity bound $S_{f,v}$ is the approximate derivative of $\Exp_{P^\prm}[f]$ with respect to a perturbation of $\prm$ in the direction of $v$. \cref{eq:dupuis239} shows that the greatest sensitivity will occur when the perturbation $v$ heavily weights likelihood parameters $\prm$ that most affect the posterior (the second term) and when earthquake parameters $f$ have the most uncertainty in the posterior (the first term). To estimate the worst-case scenario, we again assume that the perturbation $v$ is along the first singular vector of $\fim$. Such sensitivity bounds are computed for a $10\%$ relative perturbation along this direction.

To define the perturbations to the observation and likelihood distributions that we consider, we must parameterize the observation distributions selected for each historical account. To do so, we fix the structure of the choice of $\mob$ -- e.g., if the observational distribution is a normal distribution we continue to use a normal distribution -- but let the parameters, denoted by $\prm$, be the parameters characterizing that distribution, e.g., the mean and variance for a normal distribution.


\section{Results} \label{sec:application}

\subsection{Statistical summary of events}
Specific details on the geophysical insight and setting for each seismic scenario are provided in \cite{ringer2021methodological} and \cite{paskett2024tale} respectively.  We will instead only provide a summary of the resultant posterior distributions and the relative uncertainty and sensitivity for each.  Because the 1820 Sulawesi tsunami was in the vicinity of two distinct faults, \cite{paskett2024tale} sampled from two different posterior distributions, one from each fault.  Hence we report on three distinct posterior distributions.
\subsubsection*{1852 Banda Sea}
To ensure that all viable seismic events were considered, we initialized 14 MCMC chains at locations around the Banda arc with initial magnitudes of either $8.5$ or $9.0$ Mw. Additional chains were initialized at $8.0$ Mw; however, these were quickly discarded as they consistently failed to generate a sufficiently large wave to reach all of the observation points and therefore produced likelihoods of zero probability.  Each chain was initialized with the other sample parameters (depth offset etc.) set to zero. Each of the 14 chains was run for 24,000 samples, for a total of 336,000 samples. These samples were computed using the computational resources available through BYU's Office of Research Computing, consuming a total of nearly 200,000 core-hours in all.

\begin{figure}[ht]
    \centering
    \includegraphics[width=\textwidth]{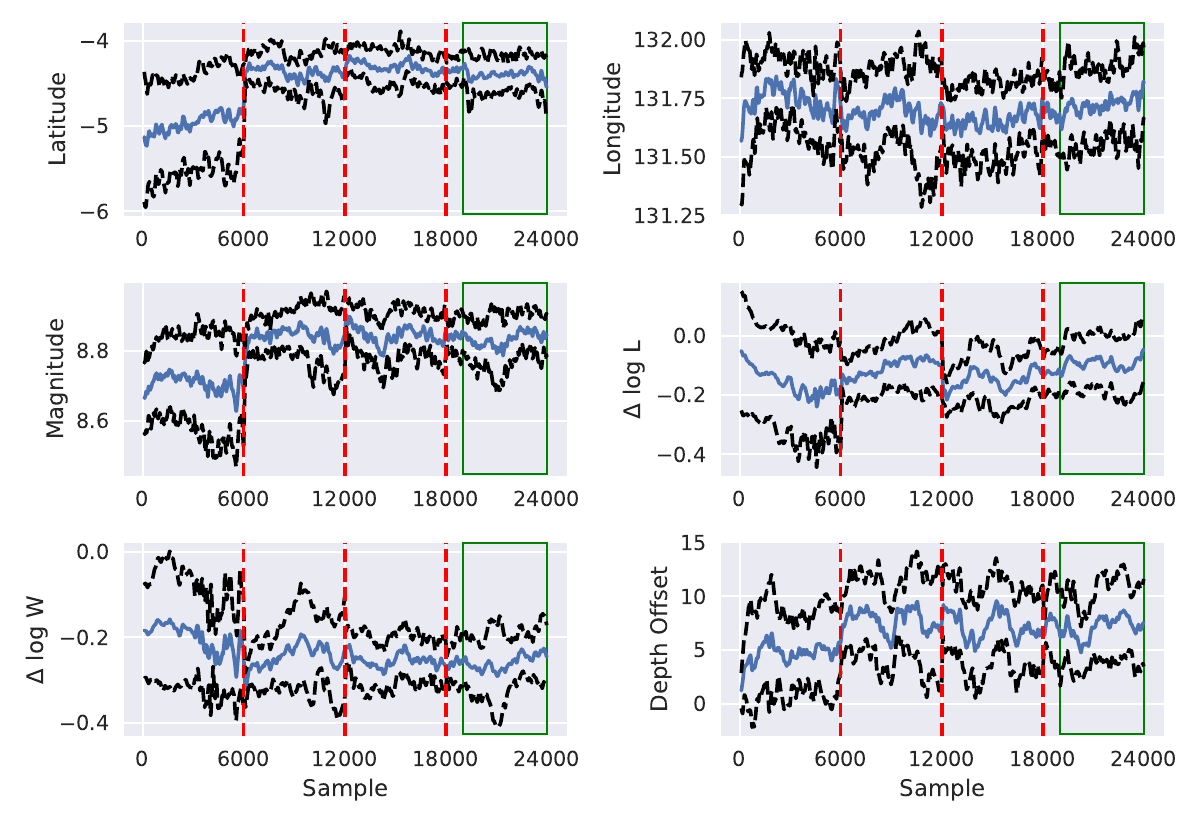}
    \caption[Sampling]{MCMC sampling by parameter. 100-sample rolling averages across all chains are shown in blue. Black lines show the rolling averages $\pm$ their standard deviations. Resampling points are marked with red lines. The green box shows samples used in the final computed posterior.}
    \label{fig:sampling}
\end{figure}

\begin{figure}[ht]
    \centering
    \includegraphics[width=0.98\textwidth]{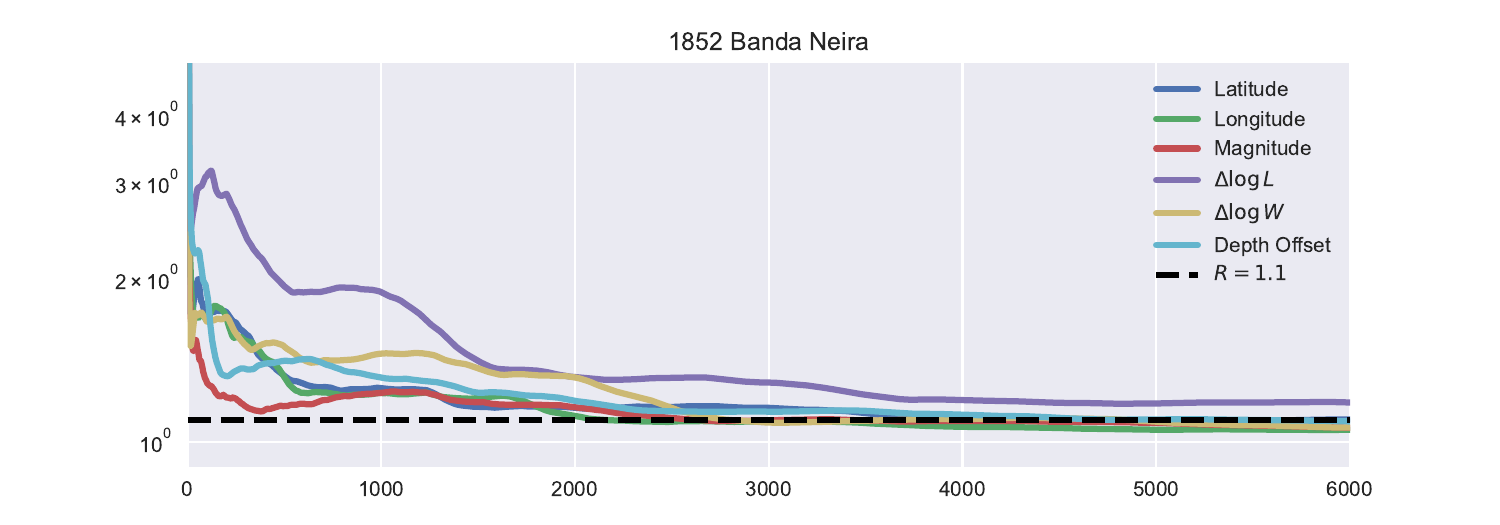}
    \caption[Gelman-Rubin Diagnostic]{The Gelman-Rubin diagnostic $\gr$ for each of the six sampling parameters for the 1852 event.  Note that computed across the final 5,000 samples the diagnostic doesn't appear to converge (upper panel), but computed across the final 10,000 samples it appears to converge very nicely (lower panel).}
    \label{fig:sampling_1852}
\end{figure}

About two thirds of the chains converged from their disparate initial conditions to a similar region in the parameter space that ultimately represented the bulk of the posterior distribution. However, the remaining third of the chains became trapped by geographic barriers in a region of parameter space with much lower posterior probability (roughly $\exp(-5)$ to $\exp(-10)$ times the probability of the samples in the first region). For this reason, after 6,000 samples we resampled the chains using importance sampling to give each chain a chance to jump to regions of higher probability. Resampling was conducted twice more at samples 12,000 and 18,000. However, the range of posterior values was much smaller at the second two resampling steps and so resampling had a less pronounced impact. Since the resampling adds a small amount of bias to the posterior once the samples are in equilibrium, these latter two resampling steps were in retrospect not warranted. To minimize their effect, we therefore use only the last 5,000 steps from each chain (assuming a 1,000 sample "burn in" after the last resampling), making a dataset of 70,000 samples from which we approximate the posterior distribution. The results of the sampling are shown in \cref{fig:sampling}; the figure shows 100-sample rolling averages across all chains (blue) plus or minus their standard deviations (black) for each parameter as well as the points at which resampling was done (red) and the samples included in the final approximate posterior (green). The figure shows the large jumps associated with the first resampling, smaller jumps at the second resampling, and almost no effect in the third resampling; the chains appear to have reached approximate equilibrium by about midway through the process so that the final 5,000 samples provide a good representation of the posterior measure as a whole. To check this, we compute the \emph{Gelman-Rubin diagnostic} $\gr$ from \cite{gelman1992inference,brooks1998general,gelman2014bayesian} for each of the six sample parameters; to ensure that the resampling does not unduly bias the results, we compute the diagnostic using the $6,000$ samples from after the last resampling step. Each of the six parameters have $\gr < 1.2$ (all but one is below $1.1$), a common criterion indicating convergence, see, e.g. \cite{brooks1998general}. A plot of $\gr$ is shown in \cref{fig:sampling_1852}.

\subsubsection*{1820 Sulawesi, Flores source}Similarly, we initialized this event with ten different chains at five distinct latitude-longitude centroid locations and with either a magnitude $8.5$ Mw or $9.0$ Mw.  Each chain was run for at least 9,000 samples after resampling with posterior probability at a burn-in of 2,000 samples, for a total of 127,690 samples.  Each sample was significantly more costly 
as the shallow Flores sea propagates the tsunami at a much slower rate so the real time simulations took nearly 2.5 months of real time simulations.
\subsubsection*{1820 Sulawesi, Walanae/Selayar source}These chains were initialized very similarly along the Walanae/Selayar fault with the same burn-in, but for a total of 104,970 total samples due to a slightly longer runtime per sample.  Even so each chain achieved a minimum of 9,000 samples and the total computation time was approximately 2.5 months.  The Gelman-Rubin diagnostic for both posterior distributions for the 1820 event is depicted in Figure 6 of \cite{paskett2024tale}.  
Using this diagnostic as a guiding criterion, it appears that the sampling procedure has adequately mixed.

\subsection{Structure of the posterior distributions}
To verify that the posterior distribution for each event was adequately resolved, we approximated the Hessian near the MAP point.

\subsubsection*{1852 Banda Sea}
\begin{figure}[h]
    \centering
    \begin{subfigure}{0.472\textwidth}
	\includegraphics[width=\textwidth]{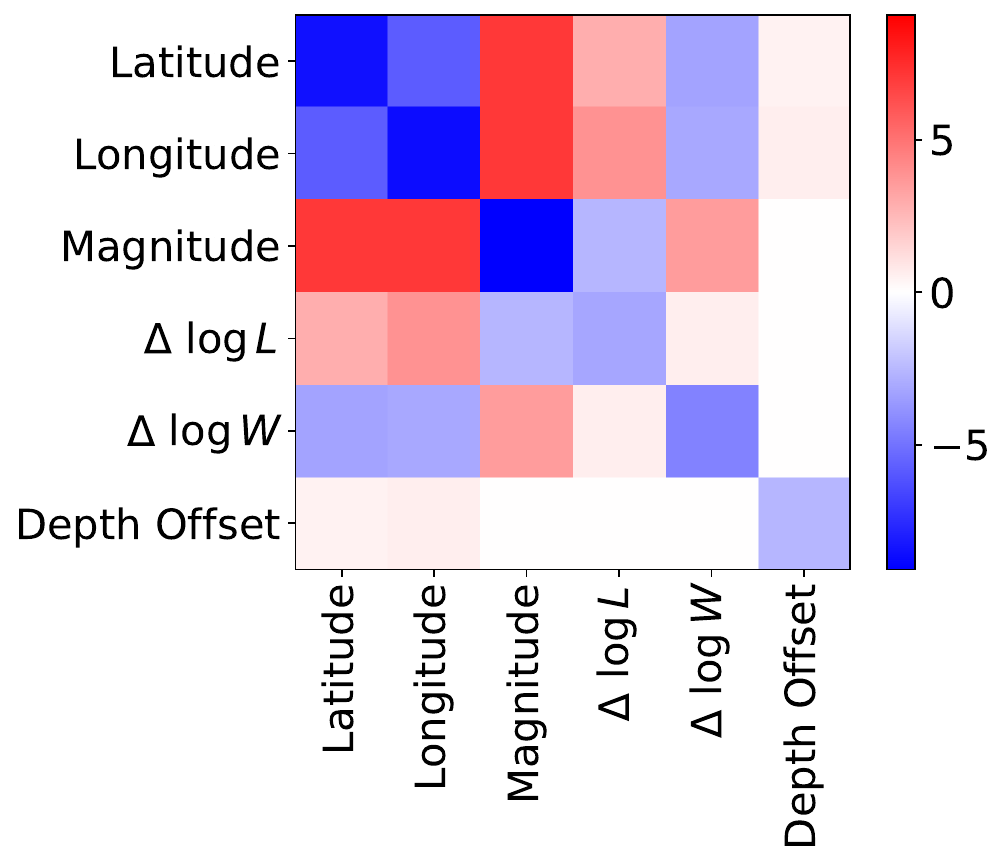}
    \end{subfigure}
    \quad
    \begin{subfigure}{0.491\textwidth}
	\includegraphics[width=\textwidth]{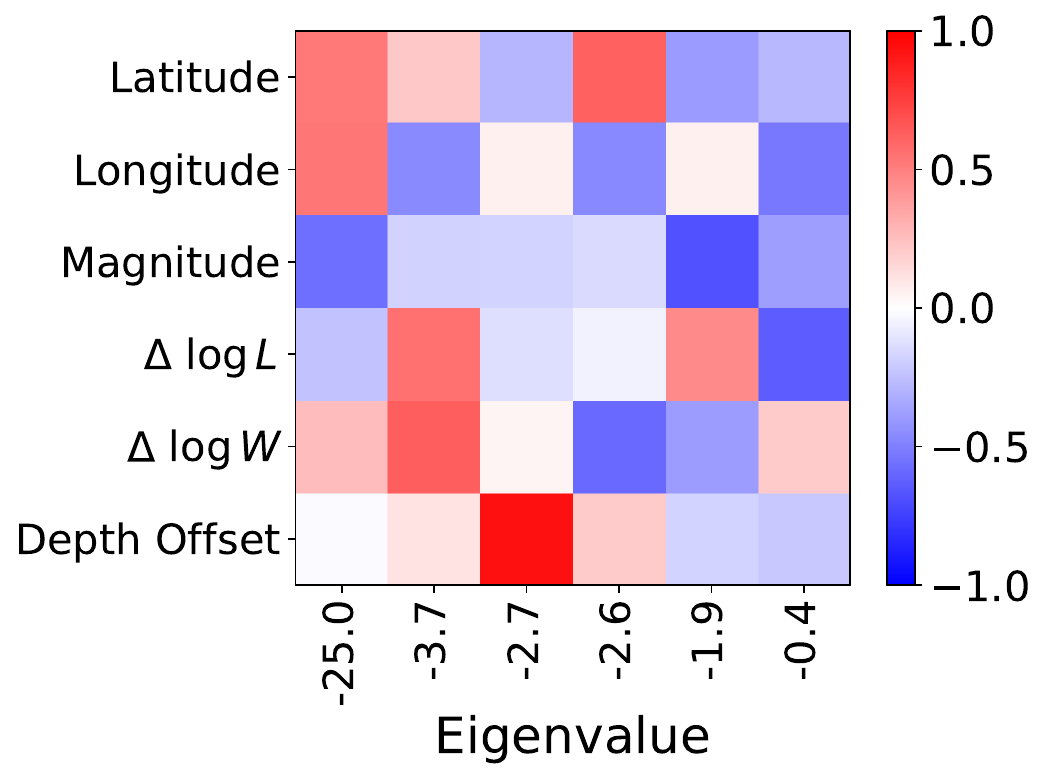}
	\vspace{0.35cm}
    \end{subfigure}
    \caption[Estimated Hessian of the log-posterior]{Estimated Hessian of the non-normalized log-posterior (left) and the corresponding eigenvectors (right) for the posterior distribution of the Banda 1852 event. The eigenvectors are presented as columns of the matrix on the right, with the corresponding eigenvalues listed on the horizontal axis. The eigenvalues indicate the relative sensitivity in each principle direction, i.e. the dominant eigenvector is approximately $25/3.7 \approx 6.75$ times more sensitive than the second most dominant eigenvector.}
    \label{fig:post:hessian}
\end{figure}

Figure~\ref{fig:post:hessian} shows the Hessian of the non-normalized log-posterior approximated at the maximum a posteriori (MAP) point and the eigenvectors of the Hessian for the Banda 1852 posterior.
Although the Hessian matrix doesn't provide the absolute magnitude of the curvature near the MAP, it does demonstrate the relative curvature in different directions.

\subsubsection*{1820 Sulawesi}

\begin{figure}[h]
    \centering
    \begin{subfigure}{0.472\textwidth}
	\includegraphics[width=\textwidth]{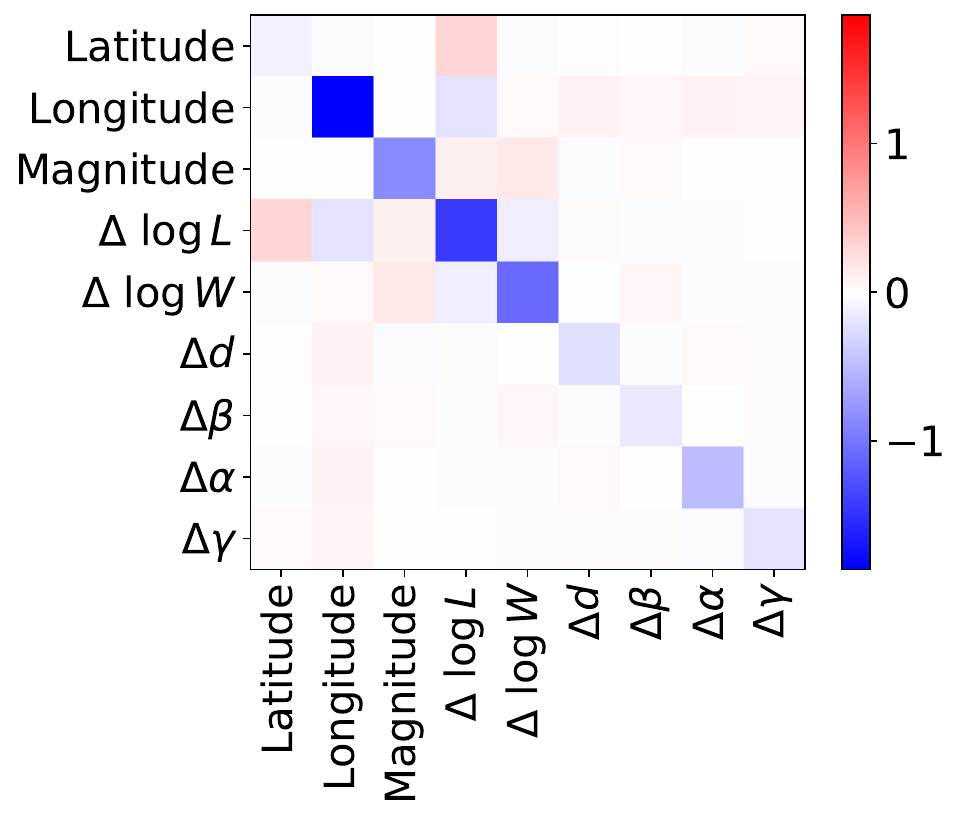}
    \end{subfigure}
    \quad
    \begin{subfigure}{0.491\textwidth}
	\includegraphics[width=\textwidth]{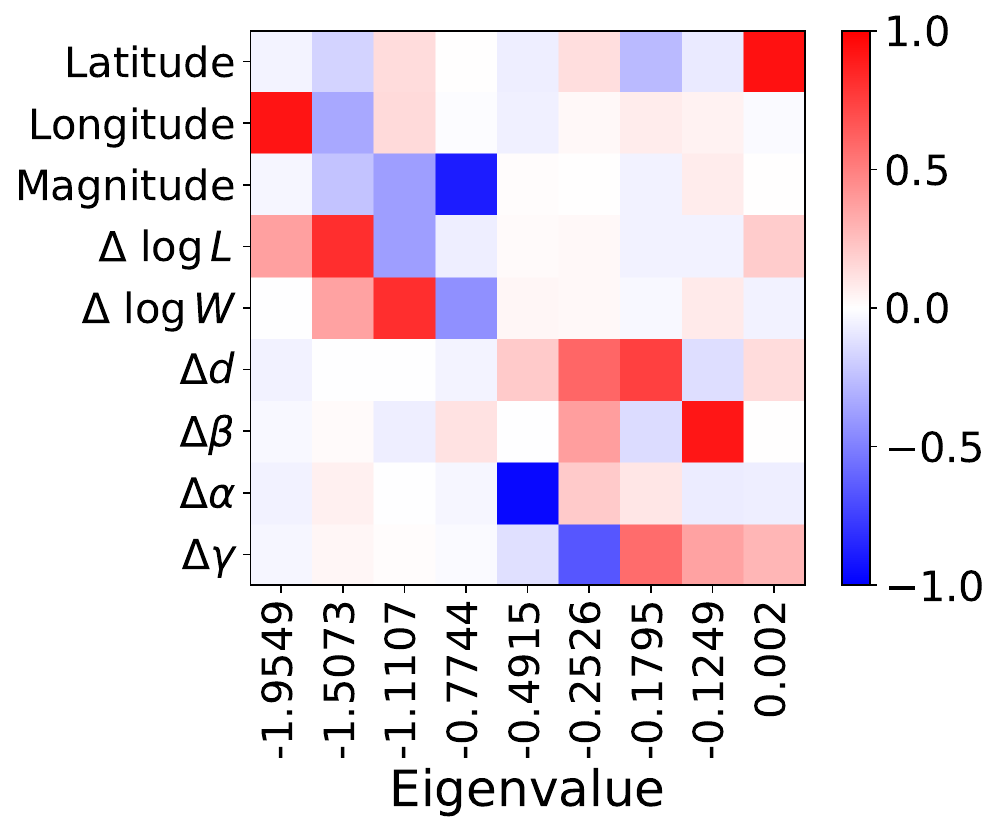}
	\vspace{-0.3cm}
    \end{subfigure}\\
    \begin{subfigure}{0.472\textwidth}
	\includegraphics[width=\textwidth]{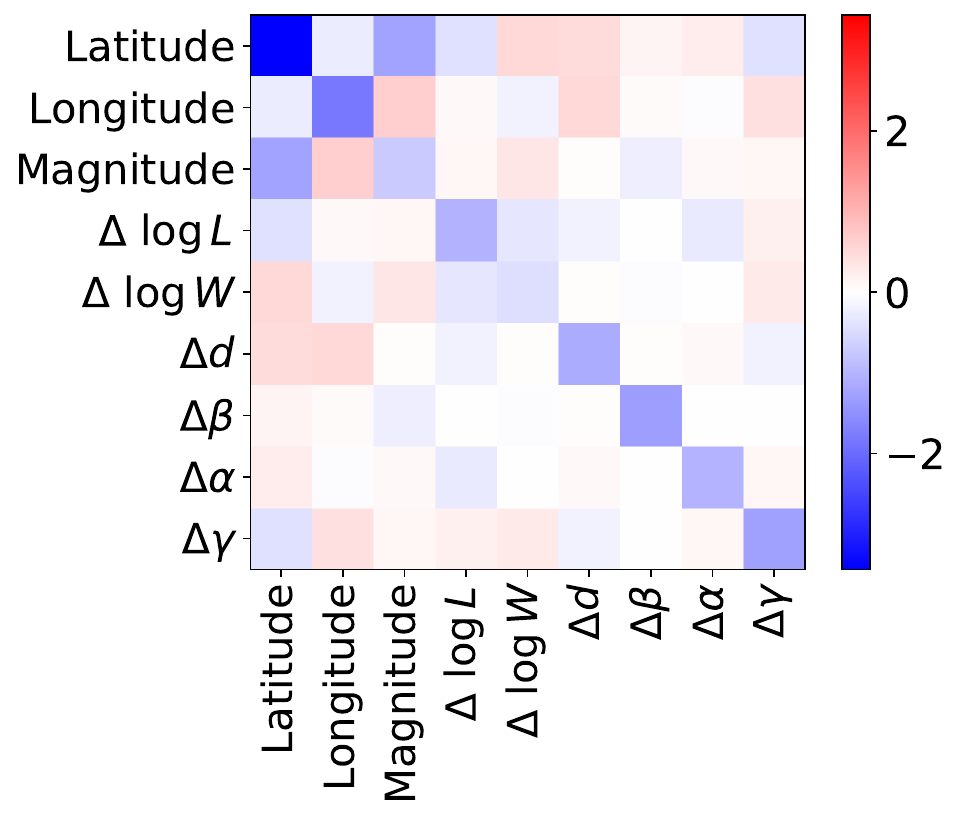}
    \end{subfigure}
    \quad\begin{subfigure}{0.491\textwidth}
	\includegraphics[width=\textwidth]{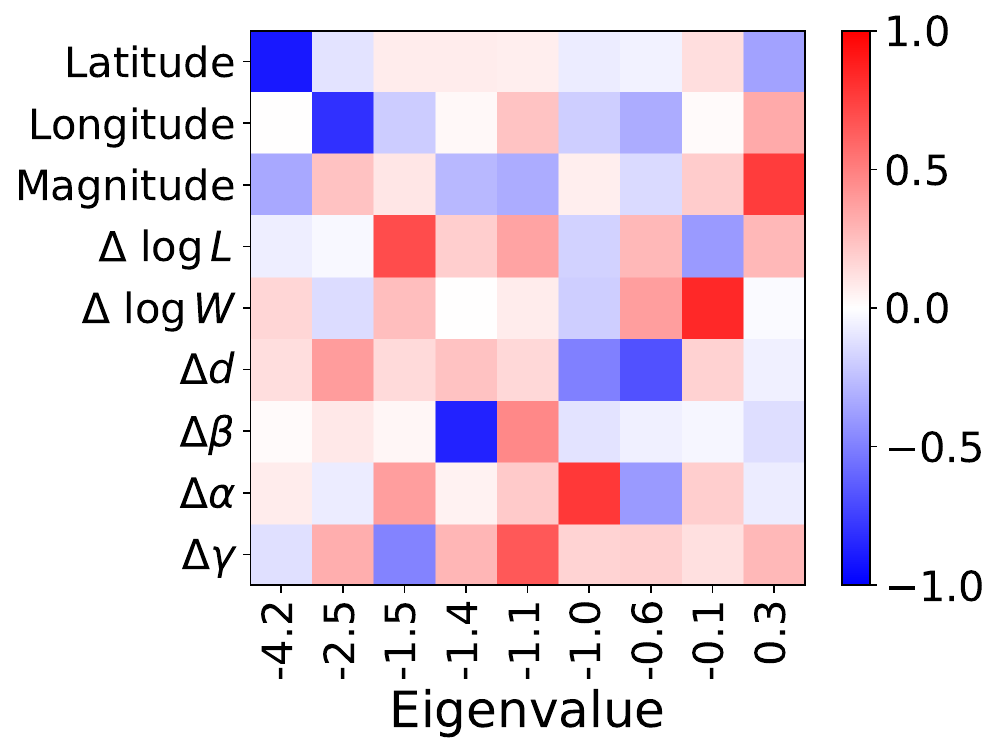}
	\vspace{0.3cm}
    \end{subfigure}
    \caption[Estimated Hessian of the log-posterior]{Estimated Hessian of the non-normalized log-posterior (left) and the corresponding eigenvectors (right) for the posterior distribution of the 1820 Sulawesi, Flores source (top) and Walanae/Selayar source (bottom). The eigenvectors are presented as columns of the matrix on the right, with the corresponding eigenvalues listed on the horizontal axis. The eigenvalues indicate the relative sensitivity in each principle direction.}
    \label{fig:post:hessian:1820}
\end{figure}

Figure \ref{fig:post:hessian:1820} presents the same calculation as Figure \ref{fig:post:hessian}, but now for the 1820 Sulawesi event.  There are several noteworthy features of this calculation. Before discussing these issues, we note that the dimension of the posterior for the 1820 Sulawesi event is larger because we allowed for variations in the depth, i.e. depth offset ($\Delta d$), dip offset ($\Delta\beta$), strike offset ($\Delta \alpha$), and rake offset ($\Delta \gamma$).  Hence the Hessian for the 1820 Sulawesi event is an $9\times 9$ matrix rather than the $6\times 6$ object identified in Figure \ref{fig:post:hessian} for the 1852 Banda Sea event.

\subsection{Error bounds and sensitivity of the posterior} \label{sec:res:errorbounds}

\cref{tab:perturb_banda} and \cref{tab:perturb_1820} show our first set of sensitivity estimates derived using the methodology described in \cref{sec:errorbounds}. First, for each $\prm$ parameter we list the associated Fisher information (the diagonal element of the FIM).  The FIM is here calculated with respect to the parameters that describe the observational probability distributions, i.e. we are considering the sensitivity of the posterior with respect to the likelihood as specified through the observational probability distributions, as described in \cref{sec:llh}. The definition of the FIM and a derivation of it for the posterior for this problem are given in \cref{app:fim}. Because the differences in Fisher information for absolute changes in parameter values are largely driven by units (e.g., meters for wave height vs. minutes for arrival time), the Fisher information values presented in these tables are computed for the relative change in each parameter value. 

\begin{table}[!h]
  \centering
 \caption{Observation distribution parameters for the 1852 Banda Sea event, the associated Fisher Information values, relative entropy according to \eqref{eq:dupuis235} associated with a 10\% relative perturbation, and the first (most sensitive) singular vector of the Fisher information matrix.}
 \label{tab:perturb_banda}
 \footnotesize
 \begin{tabular}{lllllrrrr}
\hline
        Name & Observation & Distribution & Parameter ($\prm$) & Value &     FI & $\RE~10\%$ & Sing. Vec. \\
\hline
     Pulu Ai &      height &       normal &      mean &     3 &  5.934 & 0.030 &   -0.151 \\
     Pulu Ai &      height &       normal &       std &   0.8 &  2.505 & 0.013 &    0.087 \\
       Ambon &      height &       normal &      mean &   1.8 & 12.370 & 0.062 &    0.364 \\
       Ambon &      height &       normal &       std &   0.4 &  5.220 & 0.026 &    0.216 \\
 Banda Neira &     arrival &     skewnorm &      mean &    15 & 14.082 & 0.070 &    0.447 \\
 Banda Neira &     arrival &     skewnorm &       std &     5 &  1.950 & 0.010 &   -0.148 \\
 Banda Neira &     arrival &     skewnorm &         a &     2 &  1.339 & 0.007 &    0.132 \\
 Banda Neira &      height &       normal &      mean &   6.5 &  7.525 & 0.038 &   -0.014 \\
 Banda Neira &      height &       normal &       std &   1.5 &  0.884 & 0.004 &   -0.006 \\
 Banda Neira &  inundation &       normal &      mean &   185 &  2.663 & 0.013 &   -0.010 \\
 Banda Neira &  inundation &       normal &       std &    65 &  0.272 & 0.001 &   -0.006 \\
        Buru &      height &          chi &        mu &   0.5 &  0.006 & 0.000 &    0.009 \\
        Buru &      height &          chi &     sigma &   1.5 &  0.122 & 0.001 &    0.035 \\
        Buru &      height &          chi &       dof &  1.01 &  0.142 & 0.001 &    0.040 \\
     Hulaliu &      height &          chi &        mu &   0.5 &  0.001 & 0.000 &    0.000 \\
     Hulaliu &      height &          chi &     sigma &     2 &  0.003 & 0.000 &   -0.000 \\
     Hulaliu &      height &          chi &       dof &  1.01 &  0.185 & 0.001 &    0.002 \\
     Saparua &     arrival &       normal &      mean &    45 & 19.264 & 0.096 &    0.716 \\
     Saparua &     arrival &       normal &       std &     5 &  1.280 & 0.006 &   -0.163 \\
     Saparua &      height &       normal &      mean &     5 &  9.085 & 0.045 &    0.009 \\
     Saparua &      height &       normal &       std &     1 &  0.869 & 0.004 &   -0.005 \\
     Saparua &  inundation &       normal &      mean &   125 &  2.905 & 0.015 &    0.005 \\
     Saparua &  inundation &       normal &       std &    40 &  0.178 & 0.001 &   -0.003 \\
       Kulur &      height &       normal &      mean &     3 &  0.199 & 0.001 &    0.038 \\
       Kulur &      height &       normal &       std &     1 &  0.362 & 0.002 &   -0.050 \\
       Ameth &      height &       normal &      mean &     3 &  0.351 & 0.002 &    0.043 \\
       Ameth &      height &       normal &       std &     1 &  0.409 & 0.002 &   -0.046 \\
      Amahai &      height &       normal &      mean &   3.5 &  4.107 & 0.021 &    0.014 \\
      Amahai &      height &       normal &       std &     1 &  0.784 & 0.004 &   -0.001 \\
    \bottomrule
\end{tabular}
\end{table}

\begin{table}
  \centering
 \caption{Observation distribution parameters for the Sulawesi 1820 event, associated Fisher Information values, relative entropy according to \eqref{eq:dupuis235} associated with a 10\% relative perturbation, and the first (most sensitive) singular vector (denoted SV) of the Fisher information matrix. Note that all observation distributions were normal distributions.}
 \label{tab:perturb_1820}
 \footnotesize
    \begin{tabular}{llllrrrrrr}
\toprule
& & & & \multicolumn{3}{c}{Flores} & \multicolumn{3}{c}{Walanae/Selayar} \\
Name & Observation & Parameter & Value &     FI & $\RE~10\%$ & SV &     FI & $\RE~10\%$ & SV \\
\midrule
Bulukumba &     arrival &       mean &    15 &  0.147 & 0.001 &   -0.062 &  3.039 & 0.015 &   -0.212 \\
Bulukumba &     arrival &        std &    10 &  2.723 & 0.014 &   -0.263 &  1.017 & 0.005 &    0.026 \\
Bulukumba &      height &       mean &    18 &  2.046 & 0.010 &    0.494 &  1.716 & 0.009 &    0.398 \\
Bulukumba &      height &        std &     5 &  3.305 & 0.017 &   -0.624 &  2.242 & 0.011 &   -0.446 \\
  Sumenep &     arrival &       mean &   240 &  2.409 & 0.012 &    0.093 &  0.266 & 0.001 &   -0.049 \\
  Sumenep &     arrival &        std &    45 &  1.108 & 0.006 &   -0.065 &  0.047 & 0.000 &    0.021 \\
  Sumenep &      height &       mean &   1.5 &  0.481 & 0.002 &    0.132 &  0.125 & 0.001 &    0.084 \\
  Sumenep &      height &        std &     1 &  0.072 & 0.000 &   -0.007 &  0.074 & 0.000 &   -0.063 \\
Nipa-Nipa &      height &       mean &     3 &  1.044 & 0.005 &    0.349 &  1.342 & 0.007 &    0.365 \\
Nipa-Nipa &      height &        std &     2 &  1.047 & 0.005 &    0.267 &  4.421 & 0.022 &    0.660 \\
     Bima &      height &       mean &    10 &  1.253 & 0.006 &    0.191 &  0.326 & 0.002 &    0.066 \\
     Bima &      height &        std &     4 &  1.088 & 0.005 &   -0.183 &  0.712 & 0.004 &   -0.097 \\
\bottomrule
\end{tabular}
\end{table}

Second, we present the relative entropy (Kullback–Leibler divergence) $\RE$ associated with a $10\%$ shift in each parameter computed from \eqref{eq:dupuis235}. That is, for $P^\prm$ we use the original posterior distribution; for the perturbation $v$ for the $i^{\text{th}}$ row in each table we set $v_i = 0.1 \prm_i$ and $v_j = 0, j \ne i$.   As shown in \cref{tab:perturb_banda}, this measure indicates that the most sensitive parameter in the observational probabilities (and hence the likelihood) for the 1852 event is the specified mean for the arrival time at Saparua (value of $0.096$), followed closely by the mean for the arrival time at Banda Neira (value of $0.07$) and the specified mean for the wave height at Ambon (value of $0.068$).  

The relative entropy for a $10\%$ perturbation for the 1820 event, shown in \cref{tab:perturb_1820}, is a bit more mixed and dependent on the choice of source.  For the Flores source, the dominant sensitivity according to this metric arises from the specification of the standard deviation for the wave height observation at Bulukumba (value of $0.017$) followed by the standard deviation of the arrival time at Bulukumba (value of $0.014$), the specified mean of the arrival time at Sumenep (value of $0.012$) and the specified mean of the wave height at Bulukumba (value of $0.01$).  

The last column in \cref{tab:perturb_banda} and \cref{tab:perturb_1820} lists the first singular vector of the FIM, which is the combination of perturbations of the observation parameters that produce the largest relative entropy -- effectively the ``worst-case'' perturbation, i.e. this is yet another indicator how changes in the observational probability parameters would most significantly effect the structure of the posterior. The key to interpreting these quantities is to recognize that this provides a direction along which the relative entropy will increase the most, i.e. the precise numbers provided here are less importan than their relative value relative to each other.

\begin{figure}[ht]
    \centering
    \includegraphics[width=0.98\textwidth]{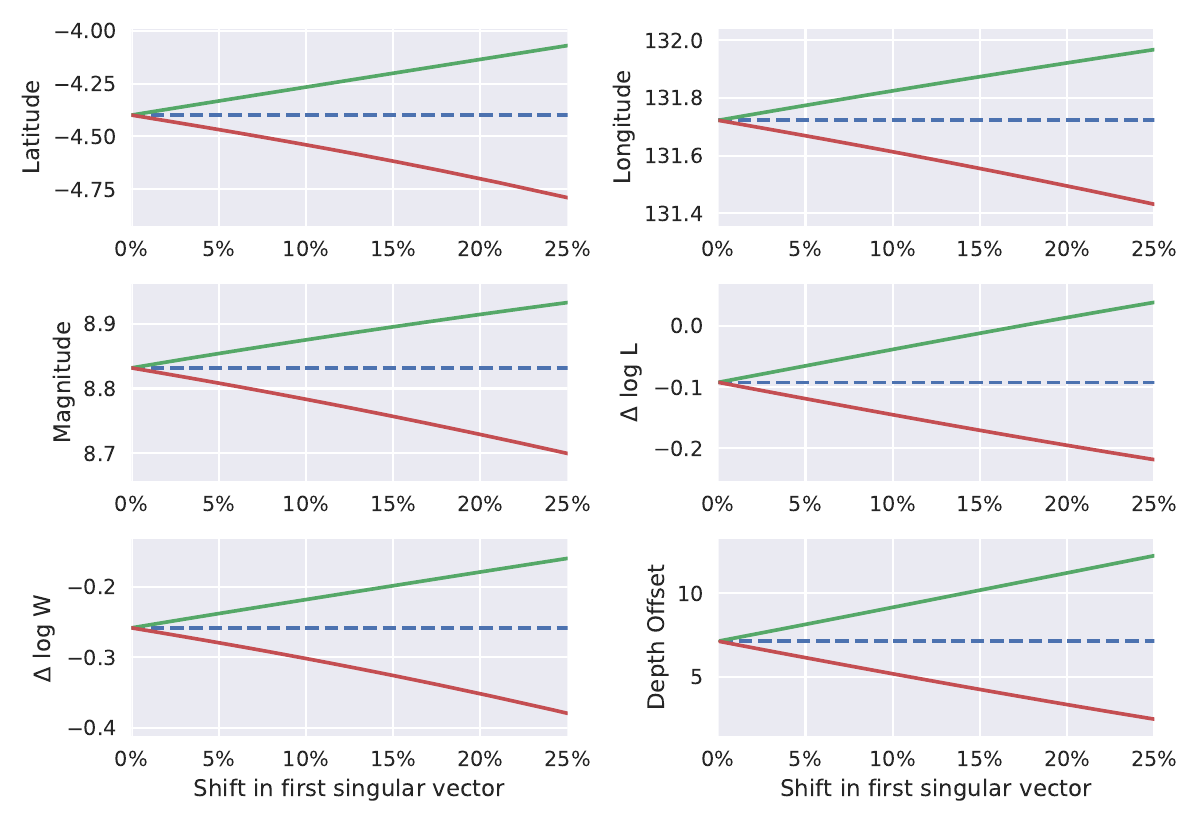}
    \caption[Bounds on Parameter Estimates]{Bounds on mean parameter values in terms of relative perturbation in the first singular vector of the Fisher information matrix (see the last column of \cref{tab:perturb_banda}) for the 1852 Banda Sea earthquake event. Upper and lower bounds are in green and red, respectively. The posterior mean is in blue.
    }
    \label{fig:banda_dupuis:211}
\end{figure}

\begin{figure}[ht]
    \centering
    \includegraphics[width=0.98\textwidth]{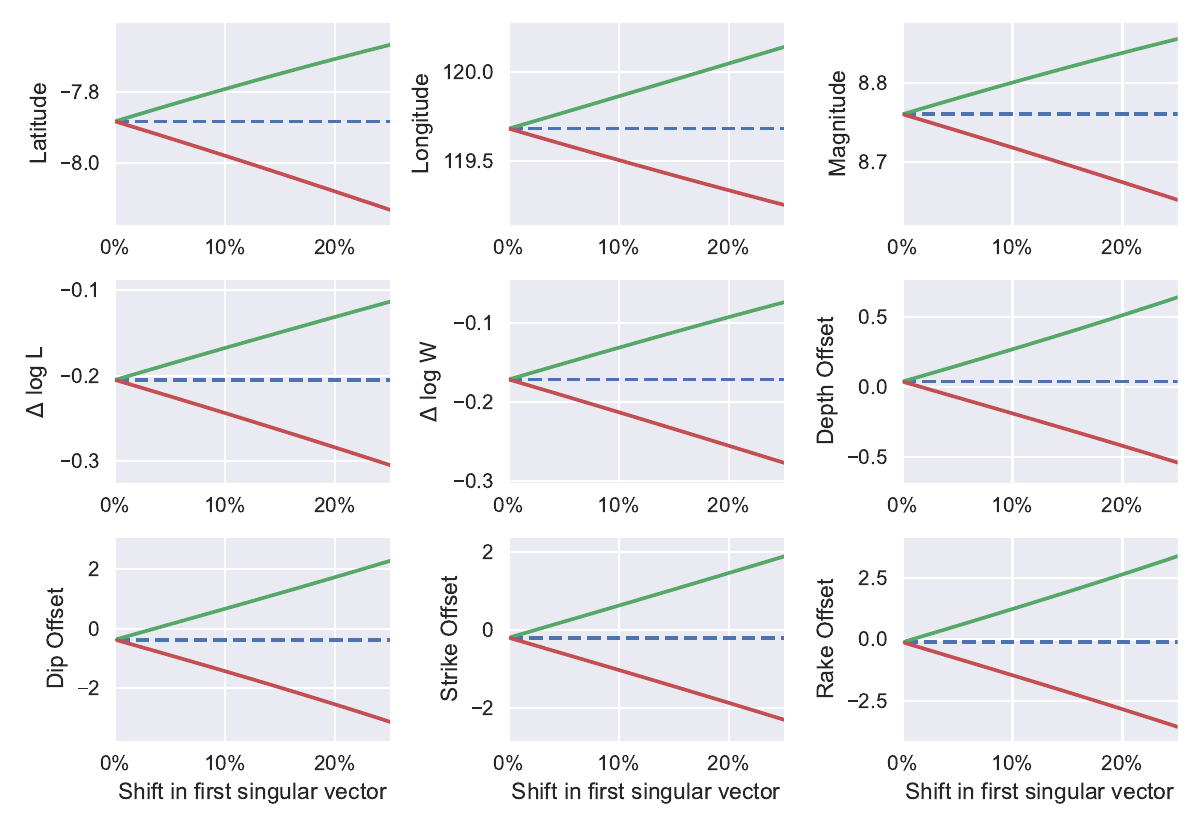}
    \caption[Bounds on Parameter Estimates]{Bounds on mean parameter values in terms of relative perturbation in the first singular vector of the Fisher information matrix (see the SV columns of \cref{tab:perturb_1820}) for the 1820 Sulawesi event for the Flores source. Upper and lower bounds are in green and red, respectively. The posterior mean is in blue.
    }
    \label{fig:flores_dupuis:211}
\end{figure}

\begin{figure}[ht]
    \centering
    \includegraphics[width=0.98\textwidth]{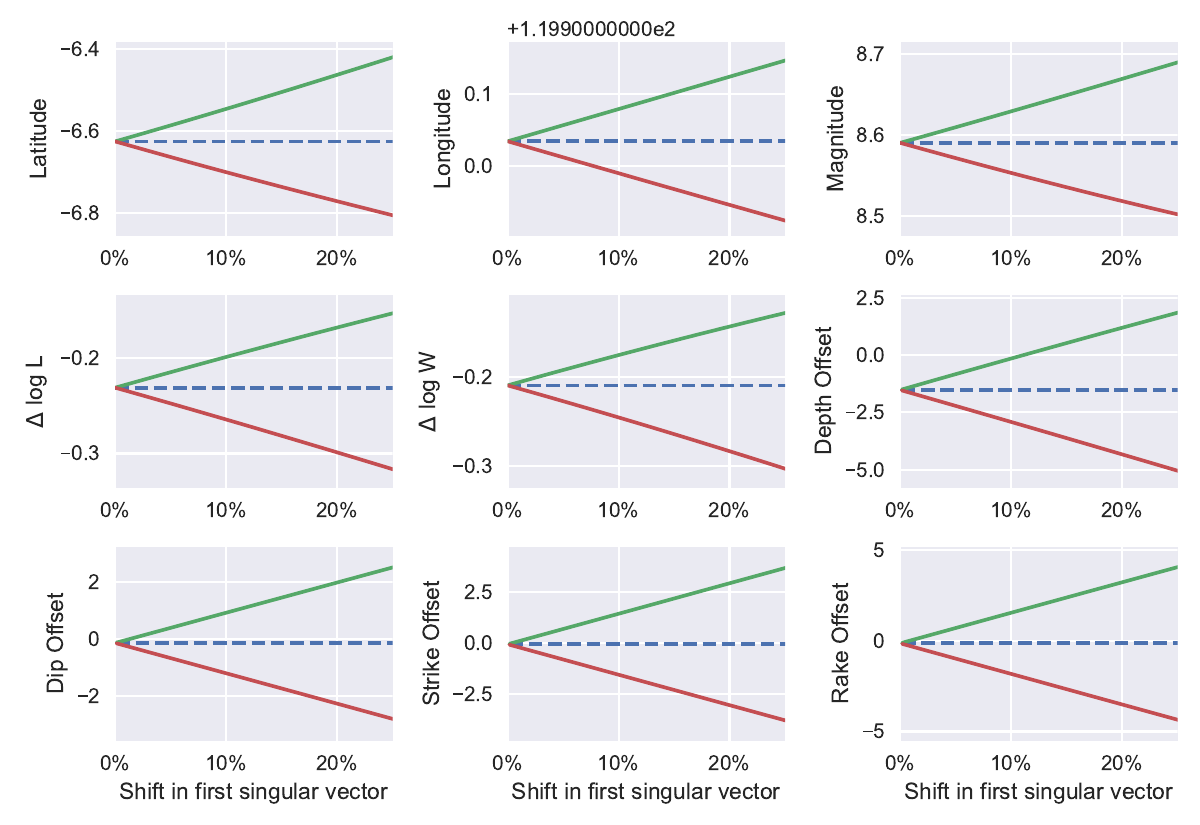}
    \caption[Bounds on Parameter Estimates]{Bounds on mean parameter values in terms of relative perturbation in the first singular vector of the Fisher information matrix (see the SV columns of \cref{tab:perturb_1820}) for the 1820 Sulawesi event with a Walanae/Selayar earthquake source. Upper and lower bounds are in green and red, respectively. The posterior mean is in blue.
    }
    \label{fig:walanae_dupuis:211}
\end{figure}

To further investigate how changes to the likelihood will affect the posterior, we projected what effect a $25\%$ shift in the parameters along the dominant singular vector will have on the mean value for the key earthquake parameters.  Even for relatively large perturbations, we get relatively narrow bounds on posterior estimates. For example, even with a 25\% perturbation in the most sensitive direction, the expected value of magnitude according to the perturbed posterior distribution would be between 8.7 and 9.0 for the 1852 Banda Sea event (see \cref{fig:banda_dupuis:211}), a very large earthquake in any case.  The 1820 Sulawesi event doesn't yield quite as strict of bounds on the estimated values for either source, but the error bars in all requisite parameters still appear reasonable (see \cref{fig:flores_dupuis:211} and \cref{fig:walanae_dupuis:211}). One caveat: As the size of the perturbation grows, the approximation in \eqref{eq:dupuis235} may break down. In this case, we refer the reader to \cref{fig:dupuis:211:re}, where the $x$-axis is in terms of relative entropy taken directly from \eqref{eq:dupuis211}.

\begin{table}[!h]
  \centering
 \caption{Posterior variance and sensitivity bound according to \eqref{eq:dupuis239} by earthquake parameter for 1852. Sensitivity bound is for relative perturbation of $10\%$ in the direction of the first (worst-case) singular vector of the Fisher Information matrix.}
 \label{tab:sensitivities_banda}
 \footnotesize
\begin{tabular}{@{}lrr@{}}
\hline
      Parameter & Variance & Sensitivity Bound \\
\hline
       Latitude &    0.066 &       0.135 \\
      Longitude &    0.040 &       0.105 \\
      Magnitude &    0.008 &       0.046 \\
 $\Delta$ log L &    0.011 &       0.054 \\
 $\Delta$ log W &    0.006 &       0.041 \\
   Depth Offset &   14.483 &       1.997 \\
\bottomrule
\end{tabular}
\end{table}

\begin{table}[!h]
    \centering
     \caption{Posterior variance and sensitivity bound according to \eqref{eq:dupuis239} by earthquake parameter for the 1820 Sulawesi event. Sensitivity bound is for relative perturbation of $10\%$ in the direction of the first (worst-case) singular vector of the Fisher Information matrix.}\label{tab:sensitivities_1820}
\begin{tabular}{lllll}
\toprule
 & \multicolumn{2}{c}{Flores} & \multicolumn{2}{c}{Walanae/Selayar} \\
 Parameter &   Variance & Sensitivity &   Variance & Sensitivity \\
\midrule
      Latitude &   0.108557 &    0.093343 &   0.108557 &    0.093343 \\
     Longitude &   0.402422 &    0.179718 &   0.402422 &    0.179718 \\
     Magnitude &   0.021577 &    0.041614 &   0.021577 &    0.041614 \\
$\Delta$ log L &   0.018109 &    0.038124 &   0.018109 &    0.038124 \\                                                                                        $\Delta$ log W &   0.020772 &    0.040831 &   0.020772 &    0.040831 \\
  Depth Offset &    0.64787 &    0.228032 &    0.64787 &    0.228032 \\
    Dip Offset &  13.573413 &    1.043748 &  13.573413 &    1.043748 \\                                                                                         Strike Offset &   8.498352 &    0.825883 &   8.498352 &    0.825883 \\
   Rake Offset &  22.272118 &    1.337002 &  22.272118 &    1.337002 \\\bottomrule
\end{tabular}
\end{table}

A final set of estimates for the sensitivity of each posterior distribution with respect to perturbations in the dominant singular vector are provided in \cref{tab:sensitivities_banda} and  \cref{tab:sensitivities_1820}.  
These estimates are computed from \eqref{eq:dupuis239}, which, as noted in \cref{sec:errorbounds}, shows that the greatest sensitivity will occur when the perturbation heavily weights likelihood parameters $\prm$ that most affect the posterior and when earthquake parameters have the most uncertainty in the posterior. To estimate the worst-case scenario, we again assume that the perturbation is a $10\%$ relative perturbation along  the first singular vector of $\fim$. Among earthquake parameters, the greatest sensitivities for the 1852 event were associated with depth offset (\cref{tab:sensitivities_banda}) because it had the widest distribution according to the posterior measure. Similar calculations are reported in \cref{tab:sensitivities_1820} for both posterior distributions of the 1820 event.


\section{Discussion}
\label{sec:discussion}

The results for the 1852 Banda Arc earthquake and tsunami show the promise of the described methodology: even though the historical accounts of the tsunami are textual in nature and therefore individually prone to much uncertainty (both aleatoric and epistemic), it nevertheless appears that taken together they can be used to determine key characteristics of the causal earthquake. The approach is similar to the ``ad hoc'' approach described in \cite{LiuHarris2014,fisherharris2016} for example, but with a more reproducible and rigorous set of assumptions, a more comprehensive coverage of possible events via automation, and a more clear characterization of uncertainty on the results. The strategy outlined in \cref{sec:llh} can readily be applied to any number of historical seismic events, but also any other problem of inverting from textual accounts similar in nature to those described in \cref{sec:data} or any other historical or other data that is similarly ``small'' -- sparse and riddled with uncertainty, so long as a reasonably believable forward model is available with parameters on which we can formulate a suitable prior distribution.

Our formal quantification of the uncertainty for this inverse problem provides justification that the 1852 Banda event was statistically adequately resolved, and also yields additional information on the potential seismic source of the 1820 event.  We discuss each of these events individually below.

\subsection*{Discussion of the 1852 Banda Sea Posterior Distribution}

We first observe that the Hessian approximation for the 1852 Banda Sea event depicted in \cref{fig:post:hessian} indicates that the posterior is adequately resolved, as all six eigenvalues are clearly negative. From this same computation we see that the steepest concavity near the MAP point follows along the latitude-longitude and magnitude.
In particular, the fastest way to decrease the posterior from the MAP point is to select a centroid to the northeast of the MAP and decrease the magnitude, which would lead to a tsunami with wave heights lower than those that match the observational likelihoods as it would place the centroid further from each observation point, and lowering the magnitude would lower the wave height as well.

The sensitive dependence on the arrival times (specifically the means of the observational probability distributions assigned to the arrival times) denoted in \cref{tab:perturb_banda} for the 1852 event is expected as we anticipate that these two arrival times are the dominant observation which triangulate the latitude-longitude location of the centroid, and since there are only two arrival time observations, each one is critical to the inference of the full posterior.  The wave height at Ambon is likely emphasized because this particular observation was very precisely specified and so the resultant observational distribution was quite narrow, i.e. a small change in the mean would result in a significantly different estimated posterior.

In terms of the actual posterior distribution, 
the data in \cref{tab:sensitivities_banda} indicates that depth offset is the least certain inferred parameter for the 1852 event, i.e. we have the least acmount of confidence on the estimated values of depth offset relative to all other components of the posterior.  On the other hand, the small sensitivity for the magnitude, $\Delta \log L$, and $\Delta \log W$ indicate that the inferred values of these three parameters representing the size of the earthquake are quite certain. Hence, we have relatively high confidence in the post-dicted magnitude of the earthquake, and the least confidence in the depth.

\subsection*{Discussion of the 1820 South Sulawesi Earthquake}
Turning to the 1820 event, the true MAP point (in the sense of the maximal posteriori probability) for the Flores posterior falls near the boundary of the distribution because the preferred geographic location is on the edge of the allowable region prescribed by the prior, i.e. our choice of the prior necessarily cuts the distribution off abruptly, but the likelihood prefers the centroid to fall on the edge of this cutoff.  This is problematic for the approximation of the Hessian displayed in \cref{fig:post:hessian:1820} because there are not a sufficient number of sampled points near this MAP point that allow for an accurate approximation of the Hessian.  Instead, we defer to a secondary MAP point that is in the highest density sampled region of the posterior for the calculation presented in the top plots of Figure \ref{fig:post:hessian:1820}.  Hence, this calculation is really an evaluation of the concavity of the posterior in the region of the densest samples which is actually not the same as the true MAP point for this distribution.

The Hessians of both the Flores and Walanae/Selayar posterior distributions are diagonally dominated with a negative correlation along the diagonal indicating that the different variables are relatively uncorrelated. The dominant eigenvector for the Flores posterior indicates that the easiest route to diminishing the posterior probability is to increase the longitude, i.e. move the centroid of the earthquake source to the east.  This would increase the arrival time of the wave at Buru, but also pushes the centroid outside the established prior distribution for the Flores fault and will decrease the likelihood corresponding to the observation at Bima.  On the other hand the Walanae/Selayar posterior will optimally decrease by moving the earthquake centroid to the south and decreasing the magnitude which will yield longer arrival times and wave heights at Bulukumba which it appears is a critical observation in the construction of the posterior.

Both Hessians for the 1820 event have one eigenvalue that is positive, even though the magnitude of the positive eigenvalue is small in both cases relative to the dominant negative eigenvalue.  This is likely due to a combination of numerical error in the Hessian approximation, and a result of the peculiar nature of the 
prior distributions used to build both of these posteriors.  In particular, as noted above and in \cite{paskett2024tale} both the Flores and Walanae/Selayar posterior distributions are concentrated on the edge of the prior.  The Flores posterior is focused further north than the prior prefers and indeed is on the wrong side of the fault, whereas the Walanae/Selayar posterior is focused on the very southern edge of the fault, once again at the edge of the prior.  In particular, the eigenvector corresponding to a positive eigenvalue in the Flores case indicates that an increase in latitude would yield a higher probability, but as the Flores thrust runs primarily east-west, such a movement in the earthquake centroid is not only prohibited by the prior, but is completely unphysical.  For the Walanae/Selayar posterior, the eigenvector corresponding to the positive eigenvalue indicates that increasing the magnitude of the earthquake, and marginally moving southeast will increase the probability, but both of these changes are prohibited by the prior distribution and indeed, are not physically feasible earthquake parameters for the Walanae/Selayar fault.  Both of these observations indicate that the most likely source of the 1820 Sulawesi tsunami was on neither the Flores thrust nor the Walanae/Selayar faults alone, but may have originated on the previously undiscovered Kalatoa (see \cite{simanjuntak2023spatial}) fault and then triggered somewhere between the two hypothesized sources.  Another possibility is the near-simultaneous rupture of a combination of all three faults.  This hypothesis arises because the dominant eigenvector of the Hessian for both faults points toward the Kalatoa fault which is beyond the defined boundary of the prior for each possibility as reported here and in \cite{paskett2024tale}.

The Walanae/Selayar source has a relatively high sensitivity to the specified standard deviation of the arrival time at Bulukumba (value of $0.015$ as specified in \cref{tab:perturb_1820}) and the specified standard deviation of the wave height at Bulukumba (value of $0.011$), but seems to be most sensitive to the specified standard deviation of the wave height at Nipa-Nipa (value of $0.022$).  This may be because although the Walanae/Selayar source is a closer match to the observations in Bulukumba, it is still not very close to the historical record, with a far lower wave height and an indeterminate arrival time.  A possible interpretation for the higher sensitivity for the observation at Nipa-Nipa is that Nipa-Nipa and Bulukumba are not geographically very far apart yet the historical record indicated a vastly different wave height between the two locations which does not appear to be supported by the simulations reported here.  The Walanae/Selayar source does not appear to be sensitive to the recorded arrival time at Sumenep, indicating that the posterior seems to have adequately balanced the differences between the prior and likelihood for this observation unlike the case of the Flores source for this event.
As shown in \cref{tab:perturb_1820} the most sensitive parameter for the Flores source is the standard deviation of the wave height at Bulukumba, which likely indicates that the specified source can't adequately capture the variation in the wave heights at this location without significantly modifying the observational probability distribution here.

The most significant changes to the posterior distribution for both sources appears to be making smaller waves at Bulukumba less plausible (push the mean up and the standard deviation down) and making large waves at Nipa-Nipa more plausible (increase both mean and standard deviation).  This is likely because these two observation points are geographically close, yet the recorded wave height at Nipa-Nipa is much smaller than that recorded at Bulukumba, i.e. the most likely change in the observational probabilities to match the simulations would be to put these two observations more in line with each other.

Variance and sensitivity bounds for the 1820 event (\cref{tab:sensitivities_1820}) indicate that the inference did not provide much information about the value of the dip, strike, and rake offset parameters, whereas parameters related to earthquake size and location were more constrained and robust to changes in the likelihood parameters.  Essentially the posterior distribution provides a reasonable post-diction on earthquake magnitude and centroid location whereas the other geometric aspects of the fault are far less constrained by the historical tsunamigenic record.

\section{Conclusions} \label{sec:conclusions}
We have presented an approach to Bayesian inference that incorporates both aleatoric and epistemic uncertainty into the inverse problem.  We demonstrate the utility of this approach on the problem of inferring the earthquake parameters for causal earthquakes that generated historical tsunamis in the Indonesian archipelago.  We then use several tools from probability and numerical analysis to provide estimates on the uncertainty that arises from the solution to this inverse problem.  In doing so, we have clear evidence that the posterior distribution for the 1852 Banda Sea earthquake as reported in \cite{ringer2021methodological} is fully resolved.  We also found further indication through these methods that the 1820 tsunami in the Flores Sea was most probably caused by an earthquake source along the newly discovered Kalatoa fault, or perhaps even according to a multi-rupture scenario akin to the 2023 dual rupture in Turkey (see \cite{mai2023destructive} for example) or the 2016 Kaikoura rupture (see \cite{cesca2017complex} for example) in New Zealand.  All of these results emphasize the need to include such statistical, probabilistic, and numerical estimators/bounds in any discussion of such problems where both the aleatoric and epistemic uncertainty are high.

\begin{appendix}

\section{Derivation of the Fisher Information Matrix}\label{app:fim}

In this appendix we derive the Fisher Information Matrix (FIM) $\fim$ associated with the parameterization of the posterior measure given in \eqref{eq:bayes}. The FIM associated with parameter $\prm$ is given by
\begin{align} \label{eq:fim}
    \fim(P^{(\prm)}) = \int \nabla_\prm \log p^{(\prm)}(x) \left( \nabla_\prm \log p^{(\prm)}(x) \right)^T P^{(\prm)}(dx)
\end{align}
where $P^{(\prm)}$ is the posterior measure and $p^{(\prm)}$ its associated density given by (see \eqref{eq:bayes} and \eqref{eq:llh})
\begin{align*}
    \mpsp(x)dx = \frac{1}{\zp} \mprp(x)\mobp\left( \fwd(x) \right)dx.
\end{align*}
Since the focus of this paper is on modeling of historical observations via observation distributions, we will focus on the case where $\prm$ describes the observation distributions. In this case, we have
\begin{align*}
    \fim_{ij}(P^{(\prm)}) = \Cov_{P^{(\prm)}}\left[ \frac{\partial}{\partial \prm_i} \potp, \frac{\partial}{\partial \prm_j} \potp \right]
\end{align*}
where $\Cov_{P^{(\prm)}}$ is the covariance according to the posterior and $\potp$ is the negative log-likelihood given by
\begin{align*}
    \potp(x) &:= -\log \mobp\left( \fwd(x) \right).
\end{align*}
Thus, to compute the FIM, we compute the derivative of $\potp$ with respect to each observation parameter (the ``score'') and then compute the covariance of each pair of scores, which we approximate using the observations associated with the approximate posterior samples. Since the individual distributions making up $\mob$ are assumed to be independent, the derivatives can be computed separately for each observation distribution. We now consider each type of observation distribution used to construct the posterior distributions in this article.

\subsection*{Normal Distribution}
For a normal distribution with mean $\mu$ and standard deviation $\sigma$, we have
\begin{align*}
    \potp(x) 
    = \frac{1}{2\sigma^2} \left[ \fwd(x) - \mu \right]^2 + \ln \sigma + \frac{1}{2}\ln (2\pi) 
\end{align*}
Then the derivatives with respect to parameters $\mu$ and $\sigma$ are given by:
\begin{align*}
    \frac{\partial}{\partial \mu} \potp(x) 
    &= -\frac{1}{\sigma^2} \left[ \fwd(x) - \mu \right]
    \\
    \frac{\partial}{\partial \sigma} \potp(x) 
    &= -\frac{1}{\sigma^3} \left[ \fwd(x) - \mu \right]^2 + \frac{1}{\sigma}.
\end{align*}

\subsection*{Skew-Norm Distribution}
For a skew-normal distribution with location $\mu$, scale $\sigma$, and skew $a$, we have
\begin{align*}
    \pot(x) 
    &= \frac{1}{2}\xt(x)^2 + \ln \sigma + \frac{1}{2}\ln (2\pi) - \ln \left[ 1+\erf\left(z(x)\right) \right]
\end{align*}
where $\erf$ is the error function and $\xt,z$ are given by
\begin{align*}
    \xt(x) &:= \frac{\fwd(x)-\mu}{\sigma} 
    \quad \text{and} \quad
    z(x) :=\frac{a\xt(x)}{\sqrt{2}}.
\end{align*}
Then the derivative with respect to $z$ and $\xt$ are given by
\begin{align*}
    \frac{\partial}{\partial z}\pot(x) 
    &= -\frac{2}{\sqrt{\pi}}\frac{e^{-z(x)^2}}{1+\erf\left(z(x)\right)}
    \\
    \frac{\partial}{\partial \xt}\pot(x) &= \frac{\partial}{\partial \xt}\pot(x) + \frac{\partial \pot}{\partial z}(x)\frac{\partial z}{\partial \xt}(x) 
    = \xt(x) - \sqrt{\frac{2}{\pi}}a\frac{e^{-z(x)^2}}{1+\erf\left(z(x)\right)},
\end{align*}
so that the derivatives with respect to parameters $a$, $\mu$, and $\sigma$ are
\begin{align*}
    \frac{\partial}{\partial a}\pot(x) &= \frac{\partial \pot}{\partial z}(x)\frac{\partial z}{\partial a}(x)
    = -\sqrt{\frac{2}{\pi}}\xt(x)\frac{e^{-z(x)^2}}{1+\erf\left(z(x)\right)}
    \\
    \frac{\partial}{\partial \mu}\pot(x) &= \frac{\partial \pot}{\partial \xt}(x)\frac{\partial \xt}{\partial \mu}(x) 
    = -\frac{1}{\sigma}\left[ \xt(x) - \sqrt{\frac{2}{\pi}}a\frac{e^{-z(x)^2}}{1+\erf\left(z(x)\right)} \right]
    \\
    \frac{\partial}{\partial \sigma}\pot(x) &= \frac{\partial}{\partial \sigma}\pot(x) + \frac{\partial \pot}{\partial \xt}(x)\frac{\partial \xt}{\partial \sigma}(x)
    = \frac{1}{\sigma}\left[ 1 - \xt(x)^2 + \frac{2z(x)}{\sqrt{\pi}}\frac{e^{-z(x)^2}}{1+\erf\left(z(x)\right)} \right].
\end{align*}

\subsection*{Chi Distribution}
For the Chi distribution with location $\mu$, scale $\sigma$, and degrees of freedom $k$, we have
\begin{align*}
    \pot(x) 
    &= \frac{1}{2}\xt(x)^2 + \ln \sigma + \left(\frac{k}{2} - 1\right)\ln 2 + \ln \Gamma \left( k/2 \right) - (k-1)\ln \xt(x),
\end{align*}
where $\Gamma$ is the gamma function and $\xt$ is given by
\begin{align*}
    \xt(x) := \frac{\fwd_i(x)-\mu}{\sigma}.
\end{align*}
The derivative with respect to $\xt$ is given by
\begin{align*}
    \frac{\partial \pot}{\partial \xt} (x)
    &= \xt(x) - (k-1)\xt(x)^{-1}.
\end{align*}
Then the derivatives with respect to parameters $\mu$, $\sigma$, and $k$ are given by
\begin{align*}
    \frac{\partial \pot}{\partial \mu}(x) 
    &= \frac{\partial \pot}{\partial \xt} \frac{\partial \xt}{\partial \mu} 
    = -\frac{1}{\sigma} \left( \xt(x) - (k-1)\xt(x)^{-1} \right)
    \\
    \frac{\partial \pot}{\partial \sigma}(x) 
    &= \frac{\partial \pot}{\partial \sigma} + \frac{\partial \pot}{\partial \xt} \frac{\partial \xt}{\partial \sigma} 
    = -\frac{1}{\sigma} \left( \xt(x)^2 - k \right).
    \\
    \frac{\partial \pot}{\partial k}(x) 
    &= \frac{1}{2}\ln 2 + \frac{1}{2}\psi \left( k/2 \right) - \ln \xt(x)
\end{align*}
where $\psi$ is the digamma function.

\section{Derivation of Bounds in Terms of Relative Entropy}\label{app:dupuis211}
\begin{figure}[ht]
    \centering
    \includegraphics[width=\textwidth]{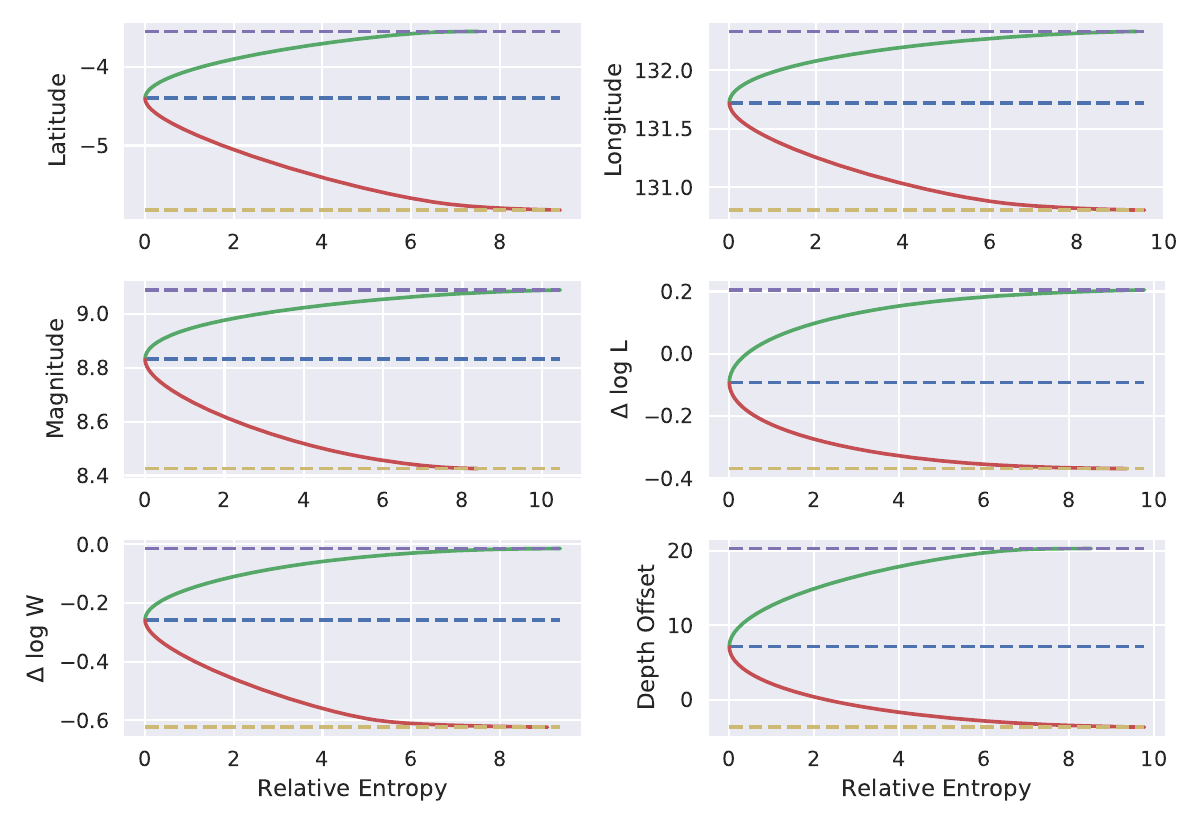}
    \caption[Bounds on Parameter Estimates]{Bounds on mean parameter values by relative entropy from computed posterior. Upper and lower bounds are in green and red, respectively. The posterior mean is in blue and the estimated uniform upper and lower bounds are in purple and yellow, respectively.}
    \label{fig:dupuis:211:re}
\end{figure}

In this section, we derive from \eqref{eq:dupuis211} more explicit bounds on $\Exp_{Q} [f]$ -- first in terms of $\RE(Q||P)$ and then independent of $\RE(Q||P)$. These bounds were used to generate \cref{fig:dupuis:211:re}, below, which shows the bounds on estimates of parameters of the 1852 Banda Arc earthquake in terms of $\RE(Q||P)$. These bounds are then combined with the estimate from \eqref{eq:dupuis235} to produce bounds in terms of relative parameter value changes.

\begin{figure}[ht]
    \centering
    \includegraphics[width=\textwidth]{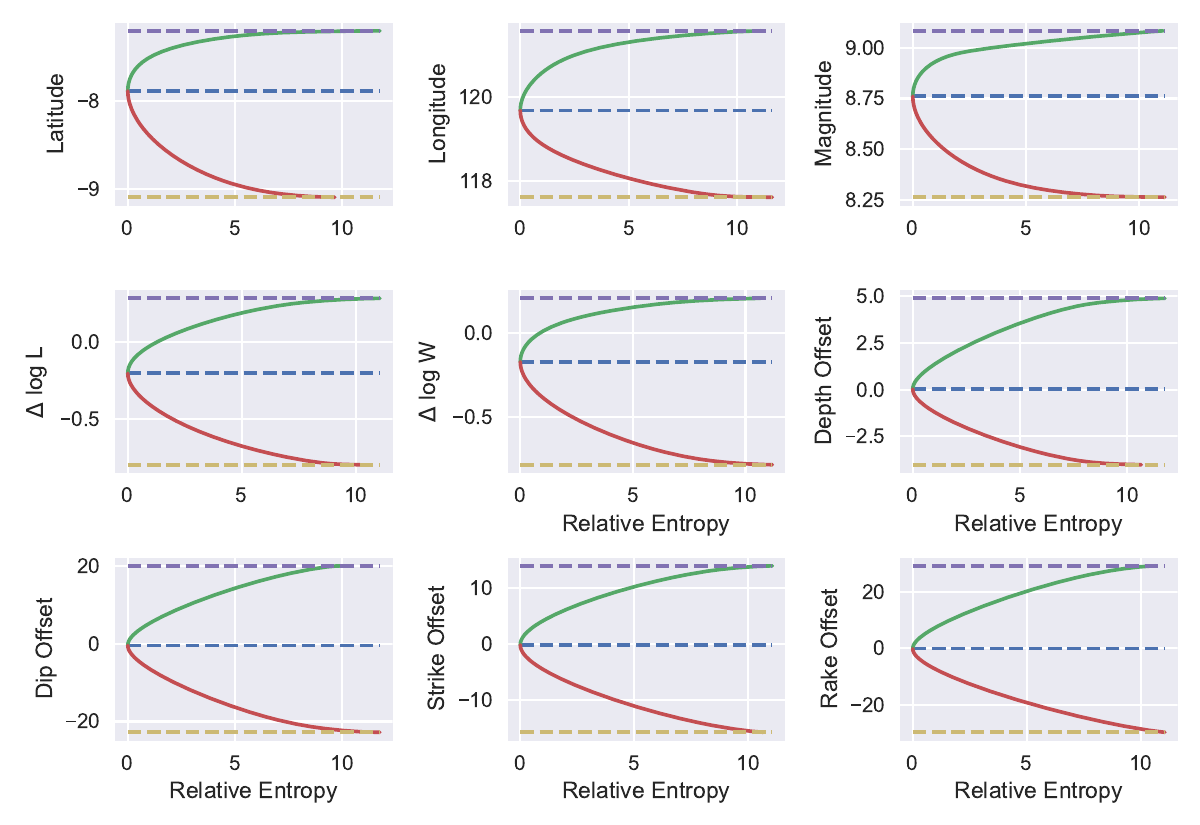}
    \caption[Bounds on Parameter Estimates]{Bounds on mean parameter values by relative entropy from computed posterior for the 1820 Flores. Upper and lower bounds are in green and red, respectively. The posterior mean is in blue and the estimated uniform upper and lower bounds are in purple and yellow, respectively.}
    \label{fig:dupuis_Flores:211:re}
\end{figure}

\begin{figure}[ht]
    \centering
    \includegraphics[width=\textwidth]{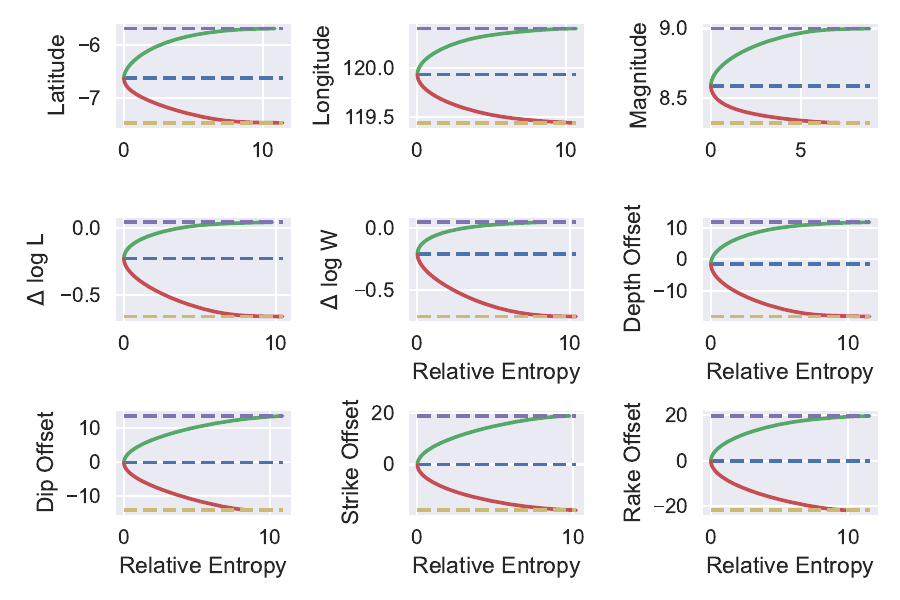}
    \caption[Bounds on Parameter Estimates]{Bounds on mean parameter values by relative entropy from computed posterior for the 1820 Walanae. Upper and lower bounds are in green and red, respectively. The posterior mean is in blue and the estimated uniform upper and lower bounds are in purple and yellow, respectively.}
    \label{fig:dupuis_Walanae:211:re}
\end{figure}

\subsection{Optimal Bound for a Given $\RE(Q||P)$}
Here we derive a relationship between the relative entropy $\RE(Q||P)$ and $c < \infty$ for which the bounds given by \eqref{eq:dupuis211} are achieved. We first assume that such a $c$ exists; the case where the supremum/infimum are achieved as $c \to \infty$ is discussed in the next subsection. Denoting $\ft = f -\Exp_{P} [f]$ and differentiating the right hand side of \eqref{eq:dupuis211} with respect to $c$ and setting the corresponding value equal to zero yields that the infimum of the upper bound is achieved when $c$ satisfies
\begin{align} \label{eq:dupuis211:Rub}
    \RE (Q || P ) &= c\frac{\Exp_{P}\left[\ft e^{c\ft}\right]}{\Exp_{P}\left[e^{c\ft}\right]} - \log \Exp_{P}\left[e^{c\ft}\right] 
    = c\frac{E_2(f,c)}{E_1(f,c)} - \log E_1(f,c)
\end{align}
and similarly for the lower bound $c$ must satisfy
\begin{align} \label{eq:dupuis211:Rlb}
    \RE (Q || P ) &= -c \frac{\Exp_{P}\left[\ft e^{-c\ft}\right]}{\Exp_{P}\left[e^{-c\ft}\right]} - \log \Exp_{P}\left[e^{-c\ft}\right] 
    =  -c \frac{E_2(f,-c)}{E_1(f,-c)} - \log E_1(f,-c)
\end{align}
where to simplify notation we have defined
\begin{align*}
    E_1(f,c) = \Exp_{P}\left[e^{c\ft}\right] 
    \quad \text{and} \quad
    E_2(f,c) = \Exp_{P}\left[\ft e^{c\ft}\right].
\end{align*}
Denoting the $c$ achieving these upper and lower bounds by $c_{+}$ and $c_{-}$, respectively, and plugging these values back into \eqref{eq:dupuis211} yields the bounds
\begin{align} \label{eq:dupuis211:R}
    \frac{E_2(f,-c_{-})}{E_1(f,-c_{-})}
    \le \Exp_{Q}[f] - \Exp_{P}[f] 
    \le \frac{E_2(f,c_{+})}{E_1(f,c_{+})}.
\end{align}
It is not clear how to invert \eqref{eq:dupuis211:Rub} and \eqref{eq:dupuis211:Rlb} to find the optimal $c$ for a given $\RE (Q || P )$, so to generate \cref{fig:dupuis:211:re} we generate a list of $c$ values, plug them into \eqref{eq:dupuis211:Rub} to find the $\RE (Q || P )$ for which they achieve the optimal upper bound and into \eqref{eq:dupuis211:R} to compute the associated upper bounds (analogously for the lower bounds), and plot those values against each other.

\subsection{Bounds Independent of $\RE(Q || P)$}
In this section, we consider the case where the bounds in \eqref{eq:dupuis211} are achieved as $c \to \infty$, i.e., are independent of $\RE(Q || P)$. From \eqref{eq:dupuis211}, we have for any $Q \ll P$
\begin{align*}
    \Exp_{Q}[f] - \Exp_{P}[f] &\le \sup_{c>0} \frac{1}{c} \log \Exp_{P}\left[e^{c(f -\Exp_{P} [f])}\right] + \frac{1}{c} \RE (Q || P ).
\end{align*}
Then clearly if
\begin{align}\label{eq:ub:finite}
    \ub := \lim_{c\to\infty} \frac{1}{c} \log \Exp_{P}\left[e^{c(f -\Exp_{P} [f])}\right] < \infty,
\end{align}
then for any $Q \ll P$ we have
\begin{align*}
    \Exp_{Q}[f] - \Exp_{P}[f] &\le \lim_{c\to\infty}\left\{ \frac{1}{c} \log \Exp_{P}\left[e^{c(f -\Exp_{P} [f])}\right] + \frac{1}{c} \RE (Q || P )\right\} = \ub.
\end{align*}
Finally, we note that since the logarithm and exponential are continuous, we have 
\begin{align*}
    \ub 
    &= \lim_{c\to\infty} \frac{1}{c} \log \Exp_{P}\left[e^{c(f -\Exp_{P} [f])}\right] 
    = \lim_{c\to\infty} \log \left( \Exp_{P}\left[e^{c(f -\Exp_{P} [f])}\right] \right)^{1/c} \\
    &= \log \lim_{c\to\infty} \left( \Exp_{P}\left[e^{c(f -\Exp_{P} [f])}\right] \right)^{1/c} 
    = \log \left\| e^{f -\Exp_{P} [f]}\right\|_{\infty} 
    = \esssup_P [f] -\Exp_{P} [f] 
\end{align*}
An analogous relationship will hold for the lower bound. Thus if $f$ is an essentially bounded random variable according to $P$, we have the following bound for any $Q \ll P$:
\begin{align*}
    \essinf_P [f] \le \Exp_{Q}[f] \le \esssup_P [f].
\end{align*}

\end{appendix}


\section*{Acknowledgements}
The authors acknowledge the Office Research Computing at BYU (\url{http://rc.byu.edu}) and Advanced Research Computing
at Virginia Tech (\url{http://www.arc.vt.edu})  for providing
computational resources and technical support that have contributed to
the results reported within this paper.  

We also thank 
G. Simpson for pointing us toward the theoretical results that ultimately yielded \cref{sec:errorbounds}; J. Guinness and R. Gramacy for helpful feedback on an early presentation of this work; as well as C. Ashcraft, A. Avery, J. Callahan, G. Carver, J. Fullwood, S. Giddens, M. Harward, R. Hilton, C. Kesler, M. H. Klein, K. Lightheart, M. Morrise, C. Noorda, A. Robertson, T. Paskett, P. Smith, D. Stewart, R. Wonnacott, and many other students at BYU who participated in the setup of the inverse problem for the events discussed here.

\section*{Funding Information}
JAK was partially supported by NSF Grant DMS-2108791. JPW was partially supported by the Simons Foundation travel grant under 586788 as well as by NSF Grants DMS-2206762 and CCF-2343286.  NEGH was partially supported by NSF Grants DMS-1816551 and DMS-2108790. AJH was supported by NIH grant K25 AI153816 and NSF grants DMS 2152774 and DMS 2236854. JPW and RH would like to thank the Office of Research and Creative Activities at BYU for supporting several of the students' efforts on this project through a Mentoring Environment Grant, as well as generous support from the College of Physical and Mathematical Sciences and the Mathematics and Geology Departments. We also acknowledge the visionary support of Geoscientists Without Borders.

\end{document}